\documentclass[]{aa}
\usepackage[varg]{txfonts}
\usepackage{xcolor}
\usepackage{placeins}
\bibpunct{(}{)}{;}{a}{}{,}

\title{Connecting Galactic and extragalactic outflows: From the Cygnus-X cluster to active galaxies}
\author{I. M. Skretas \and L. E. Kristensen}
\institute{Niels Bohr Institute \& Centre for Star and Planet Formation, Copenhagen University, {\O}ster Voldgade 5--7, 1350 Copenhagen K, Denmark; 
              \email{iaskret@gmail.com; lars.kristensen@nbi.ku.dk}}

\date{Received <date> /
Accepted <date>}
\abstract
{Molecular outflows are commonly detected originating from both protostellar and extragalactic sources. Separate studies of low-mass, isolated high-mass, and extragalactic sources reveal scaling relations connecting the force carried by an outflow and the properties of the source that drives it, as for example the mass and luminosity.
}
{The aim of this work is twofold: first, to examine the effects, if any, of clustered star formation on the protostellar outflows and their scaling relations and, second, to explore the possibility that outflows varying in scale and energetics by many orders of magnitude are consistent with being launched by the same physical processes.}
{To that end, high-angular resolution CO $J$ = 3--2 observations were used of ten high-mass protostars in the Cygnus-X molecular cloud, obtained at the SubMilliMeter Array as part of the Protostellar Interferometric Line Survey of Cygnus-X (PILS-Cygnus). From these data, the outflow force, that is the momentum ejection rate, was measured. In addition, an extended sample of protostellar and extragalactic outflow-force measurements was assembled from existing literature to allow for a direct comparison of the scaling relations of the two types of outflows.}
{Molecular outflows were detected originating from all ten sources of the PILS-Cygnus survey, and their outflow forces are found to be in close agreement with measurements from the literature.
In addition, the comparison of the protostellar and extragalactic sources reveals, with 95\% confidence, that Class 0 protostars and extragalactic sources follow the same outflow force--bolometric luminosity correlation.
}
{
The close agreement between the Cygnus-X sources and sources of similar envelope mass and bolometric luminosity suggests that clustered star formation has no significant effect on protostellar outflows.
We find a strong indication that protostellar and extragalactic outflows are consistent with having a similar launch mechanism. The existence of such a mechanism would enable the development of a single universal outflow launch model, although more observations are required in order to verify this connection.}

\begin{document}

\maketitle
\section{Introduction}

Stars form deep inside molecular clouds by accreting their surrounding matter through a circumstellar disk. This process gives rise to strong bipolar jets, which entrain the surrounding envelope material forming extended outflows \citep[e.g.,][]{bally2016}. These protostellar outflows have often been observed, mostly using low-$J$ CO line emission \citep[e.g.,][]{Bally1983,Bachiller1990,bontemps1996,maud2015II,mottram2017}, which allows for an estimate of their properties, such as outflow velocity and extent. One of the most interesting characteristics of an outflow, which can be inferred from such observations, is the momentum rate or outflow force ($F_{\text{CO}}$). The outflow force gives the rate at which the outflow injects momentum into the surrounding envelope, and it is often used to describe the strength of an outflow.
Correlations have been detected, connecting $F_\text{CO}$ to the properties of the source (e.g., luminosity and mass) of the outflow for both protostellar \citep[e.g.,][]{bontemps1996} and extragalactic \citep[e.g.,][]{lutz2020} sources. The questions that naturally arise are, whether the physical mechanisms behind these two outflow types are the same \citep{Romero2021}, whether outflows that vary in scale and energetics by many orders of magnitude can indeed be launched by the same physical processes, and how statistical methods can be used to compare and thereby infer similarities and differences between these types of outflows.

The first of the correlations regarding $F_\text{CO}$  was initially reported by \citet{bontemps1996} for a sample of low-mass,  primarily Class I protostellar sources, where they found that $F_{\text{CO}}$ scales with the mass of the surrounding envelope ($M_\text{env}$). In the same work, the observed decrease in outflow energetics between Class 0 and Class I sources is attributed to evolutionary effects, due to the dispersion of the surrounding envelope and the reported correlation between $F_{\text{CO}}$ and $M_\text{env}$.
The correlation was later expanded to the high-mass protostellar regime by \citet{beuther2002}, using a sample of 26 relatively isolated high-mass sources. The high-mass sample was later significantly strengthened by \citet{maud2015II}, who examined outflows from 99 massive young stellar objects (MYSOs) and very compact \ion{H}{ii} regions taken from the Red MSX source Survey \citep[RMS, ][]{lumsden2002,lumsden2013} sample, reporting a similar correlation. \citet{beuther2002} also reported that the correlation appears tighter between low-mass Class I sources and high-mass sources, tentatively suggesting the absence of a proper Class 0 phase in high-mass protostars.
The low-mass sample has been expanded in more recent works \citep{yildiz2015,mottram2017}, where it is also found that the commonly used low-$J$ CO observations better describe the time-averaged behavior of an outflow, while other tracers are needed in order to look into current outflow behavior (for example high-$J$ CO transitions, H$_2$O, or \ion{O}{I}).
The existence of this correlation points to a connection between M$_\text{env}$ and the mass accretion rate (\.{M}$_\text{acc}$), which in turn is connected to the outflow \citep{duarte2013,mottram2017}.

The second correlation is with the bolometric luminosity of the source ($L_\text{bol}$) \citep{cabrit1992}. In this correlation, a split between Class 0 and Class I sources has been found by \citet{bontemps1996}, with Class 0 sources systematically showing stronger outflows than Class I sources of the same $L_\text{bol}$. Similar to the $F_\text{CO}$ - $M_\text{env}$ correlation, the $F_\text{CO}$ - $L_\text{bol}$ correlation was also expanded to the high-mass regime \citep{beuther2002,maud2015II} and the low-mass sample was expanded by \citet{yildiz2015,mottram2017}. 
In this case though, \citet{maud2015II} reports a shallower slope for the high-mass sources when compared to the low-mass sources, which, if extrapolated, falls between the Class 0 and Class I sources.

Apart from protostars, galaxies are also often seen driving outflows \citep[e.g.,][]{stuber2021}, albeit on a much larger scale. The predominant mechanisms powering extragalactic outflows are believed to be accretion onto a central supermassive black hole in galaxies hosting active galactic nuclei (AGN) or a combination of stellar processes such as supernovae and star formation in the case of starbursts \cite[see, e.g.,][ for a recent in-depth review on galactic-scale outflows, the energy sources behind them and their driving mechanisms]{veilleux2020}. 
Extragalactic outflows are most commonly observed through P-Cygni profiles in OH emission lines \citep[e.g.,][]{sturm2011,gonzalez2017} or from broad line wings detected in CO line emission \citep[e.g.,][]{cicone2014,fluetsch2019} which allow for a calculation of outflow forces in a way similar to that of the protostellar outflows.
Such measurements have been carried out extensively for low-redshift sources \citep[e.g.,][]{cicone2014,gonzalez2017,fluetsch2019,lutz2020} and also more recently for higher redshift sources \citep{spilker2020b}. 
Examining the outflow forces for scaling relations, \citet{lutz2020} detected a correlation between $F_\text{CO}$ and $L_\text{bol}$ similar to the one found for protostellar ouflows.
A possible connection between protostellar and galactic-scale outflows is further emphasized in \citet{aalto2020}, where the magneto-hydrodynamic (MHD) launch mechanism suggested for the AGN-driven narrow wind and molecular jet of NGC 1377, is expected to be similar to the suggested launch mechanism of protostellar winds.

In this work, we examine 10 sources in the Cygnus-X molecular cloud. Cygnus-X is one of the most intense nearby star-forming regions, hosting a significant number of protoclusters and high-mass protostellar objects deeply embedded within massive, dense, dusty cores \citep{motte2007}, a large number of ultra-compact \ion{H}{ii} regions \citep{cyganowski2003} as well as a significant population of OB stars \citep{wright2010}, including Cyg OB2 \citep{knodlseder2000}, one of the largest known OB associations in our Galaxy. It is estimated that Cygnus-X has a total mass of $\approx 4 \times 10^6~ \text{M}_\odot$, extends over 100 pc in diameter and the distance to its various substructures ranges from 1.3 kpc for W75/N30 up to 3.3 kpc for AFGL 2591/CygX-S26 \citep{rygl2012}.

In this paper, we present CO $J$ = 3--2 and SiO $J$ = 8--7 emission from the ten sources of the PILS-Cygnus survey \citep{vanderwalt2021}, allowing for detection of outflows and measurements of outflow properties. The outline of the paper is as follows: in Sect. \ref{sec:observations} we describe the set-up used for the observations, in Sect. \ref{sec:results} we present our results which we discuss further in Sect. \ref{sec:discussion}. Finally in Sect. \ref{sec:conclusion} we present a summary and our conclusions.

\section{Observations}
\label{sec:observations}

The ten sources examined in this work \citep[CygX-N12, N30, N38, N48, N51, N53, N54, N63, S8, S26, where the nomenclature is from][see Table \ref{table:cygnusmasslum}]{motte2007} were targeted for the PILS-Cygnus survey \citep{vanderwalt2021}, a large frequency-range line survey, covering 32 GHz of continuous frequency observations in the 345 GHz atmospheric window. All the observations were carried out with the SubMillimeter Array (SMA) on Mauna Kea, Hawaii.

The sources were selected due to their location in the same molecular cloud structure and their proximity to the Cyg-OB2 association, which makes them ideal targets to examine the effects of clustered star formation on protostellar outflows. In addition, being classified as intermediate- to high-mass sources they offer a great opportunity to bridge the gap between low- and high-mass sources in protostellar outflow studies, connecting the two samples.

\begin{table*}[htbp]
\caption{Coordinates, envelope masses, and bolometric luminosities of the Cygnus-X sources.} 
\label{table:cygnusmasslum} 
\centering 
\begin{tabular}{l c c c c c} 
\hline\hline 
Source & RA (J2000) & DEC (J2000) & $M_\text{env}$ [M$_\odot$] & $L_\text{bol}$ [L$_\odot$]& Ref. \\ 
\hline
N12 & 20:36:57.68 & 42:11:30.8 & 40 & 940 & 1 \\ 
N30 & 20:38:36.45 & 42:37:33.8 & 260 & 28000 & 1 \\ 
N38 & 20:38:59.26 & 42:22:28.6 & 60 & 300 & 2,3 \\
N48 & 20:39:01.46 & 42:22:05.9 & 40 & 2100 & 1 \\
N51 & 20:39:01.97 & 42:24:59.2 & 150 & 1500 & 1 \\
N53 & 20:39:03.13 & 42:25:52.5 & 140 & 460 & 1 \\
N54 & 20:39:04.03 & 42:25:41.1 & 40 & 160 & 2,4 \\
N63 & 20:40:05.50 & 41:32:12.6 & 50 & 490 & 1 \\
S8 & 20:20:39.29 & 39:37:54.2 & 70 & 7400 & 1 \\
S26 & 20:29:24.87 & 40:11:19.3 & 350 & 244000 & 1 \\
\hline 
\end{tabular}
\tablebib{(1) \citet{pitts2021}, (2) $M_\text{env}$ from \citet{motte2007}, (3) $L_\text{bol}$ from \citet{Irene2015}, (4) $L_\text{bol}$ from \citet{kryukova2014}.}
\end{table*}

The observations for the PILS-Cygnus survey, from which the data presented in this paper are taken, were performed with the SMA using the SWARM (SMA Wideband Astronomical ROACH2 [second generation Reconfigurable Open Architecture Computing Hardware] Machine) correlator. The observations were carried out with a combination of Compact and Extended configurations with baselines from 7 to 211 meters. All ten sources were observed for approximately the same time over ten tracks, five tracks in the compact configuration and five tracks in the extended.

The compact configuration observations were performed between 27 June and 7 August, 2017, while the extended configuration observations were taken between the 20 of October and the 10 of November 2017. During this time between 6 and 8 antennas where available in the array.

For all the observations, MWC349A was used as the complex gain calibrator, the quasars 3c273, 3c454.3 and 3c84 for bandpass calibration (depending on the exact time of the observations), and Neptune, Titan, Callisto, and Uranus for the flux calibration. We refer to \citet{vanderwalt2021} for details on the observations and calibration. 

All initial observations had uniform spectral channel widths of 140 kHz, as provided by the SWARM correlator, but were then re-binned into a spectral resolution of 560 kHz ($\approx$ 0.48 km s$^{-1}$), in order to improve noise levels. All data calibrations were performed through the Common Astronomy Software Applications (CASA) v4.7 \citep{mcmullin2007}.

Subsequent cleaning of the data was also performed through CASA v4.7, using the interactive mode of the \texttt{clean} command. The manual selection of the areas to be cleaned was preferred in this case as to better trace the extended and varying CO line emission. In addition, throughout the cleaning process, natural weighting was used, by applying the Briggs weighting function with a robust parameter of 2.

\begin{table*}
\caption{Final beam sizes and $\sigma_\text{rms}$ for the cleaned CO 3--2 and SiO 8--7 data.} 
\label{table:beamsizes} 
\centering 
\begin{tabular}{l c c c c c c c c} 
\hline\hline 
 & \multicolumn{3}{c|}{\textbf{CO 3--2}} & \multicolumn{3}{c}{\textbf{SiO 8--7}} \\ 
\hline
Source & \begin{tabular}[c]{@{}c@{}}$\sigma_\text{rms}$\tablefootmark{a}\\  {[}Jy beam$^{-1}${]}\end{tabular} & \begin{tabular}[c]{@{}c@{}}Beam \\ {[}arcsec{]}\end{tabular} & \begin{tabular}[c]{@{}c@{}}PA \\ {[}deg{]}\end{tabular} & \begin{tabular}[c]{@{}c@{}}$\sigma_\text{rms}$\tablefootmark{b}\\  {[}Jy beam$^{-1}${]}\end{tabular} & \begin{tabular}[c]{@{}c@{}}Beam \\ {[}arcsec{]}\end{tabular} & \begin{tabular}[c]{@{}c@{}}PA \\ {[}deg{]}\end{tabular} \\ 
\hline
N12 & 0.15 & 1.24 $\times$ 1.08 & 85.7 & 0.09 & 1.19 $\times$ 1.04 & 70.7 \\ 
N30 & 0.28 & 1.35 $\times$ 1.19 & 75.0 & 0.10 & 1.29 $\times$ 1.14 & 56.1 \\ 
N38 & 0.19 & 1.38 $\times$ 1.19 & 71.0 & 0.08 & 1.34 $\times$ 1.18 & 91.0 \\ 
N48 & 0.16 & 1.37 $\times$ 1.13 & 76.6 & 0.09 & 1.30 $\times$ 1.11 & 77.4 \\ 
N51 & 0.18 & 1.25 $\times$ 1.05 & 83.0 & 0.10 & 1.21 $\times$ 1.03 & --85.0\phantom{--} \\ 
N53 & 0.15 & 1.19 $\times$ 1.01 & 82.1 & 0.08 & 1.14 $\times$ 0.99 & 86.8 \\ 
N54 & 0.15 & 1.21 $\times$ 1.04 & --80.9\phantom{--} & 0.08 & 1.23 $\times$ 1.07 & 88.5 \\ 
N63 & 0.17 & 1.24 $\times$ 1.11 & 88.7 & 0.09 & 1.26 $\times$ 1.12 & 82.2 \\ 
S8  & 0.21 & 1.22 $\times$ 1.12 & 87.3 & 0.08 & 1.21 $\times$ 1.11 & 87.2 \\ 
S26 & 0.21 & 1.40 $\times$ 1.25 & 77.0 & 0.11 & 1.39 $\times$ 1.24 & 76.9 \\ 
\hline 
\end{tabular}
\tablefoot{
Calculation of $\sigma_\text{rms}$ is over a single channel with channel-widths of:
\tablefoottext{a}{0.48 km s$^{-1}$ for CO emission} and \tablefoottext{b}{1.93 km s$^{-1}$ for SiO}.
}
\end{table*}

The resulting clean data have an image size of 256 $\times$ 256 pixels and a pixel size of 0\farcs2. The spectral resolution for the CO data is kept at the initial 0.48 km s$^{-1}$ but for SiO it is further re-binned to $\sim$ 1.93 km s$^{-1}$ in order to increase the signal-to-noise ratio (S/N). The rms noise ($\sigma_\text{rms}$) and the beam sizes of the final cleaned data are shown in Table \ref{table:beamsizes}.

\section{Results}
\label{sec:results}
\subsection{Outflow maps}
\label{sec:maps}
\begin{figure*}
\centering
\includegraphics[width=0.45\linewidth]{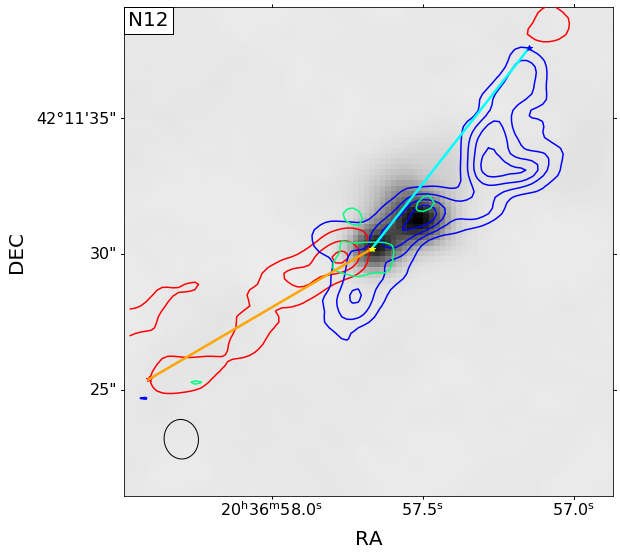}
\includegraphics[width=0.45\linewidth]{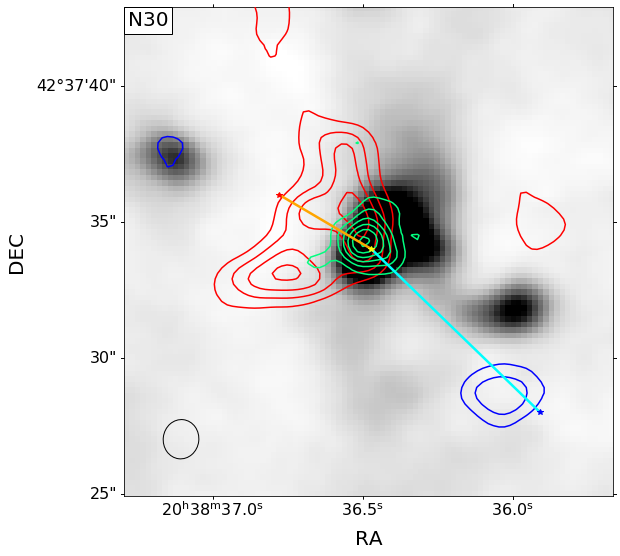}
\includegraphics[width=0.45\linewidth]{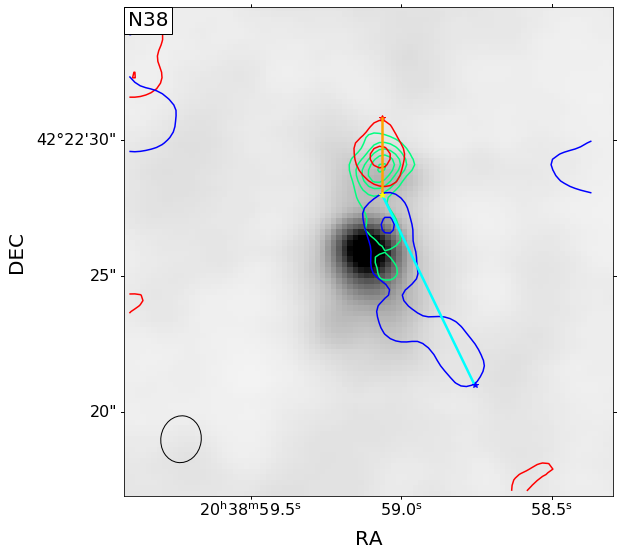}
\includegraphics[width=0.45\linewidth]{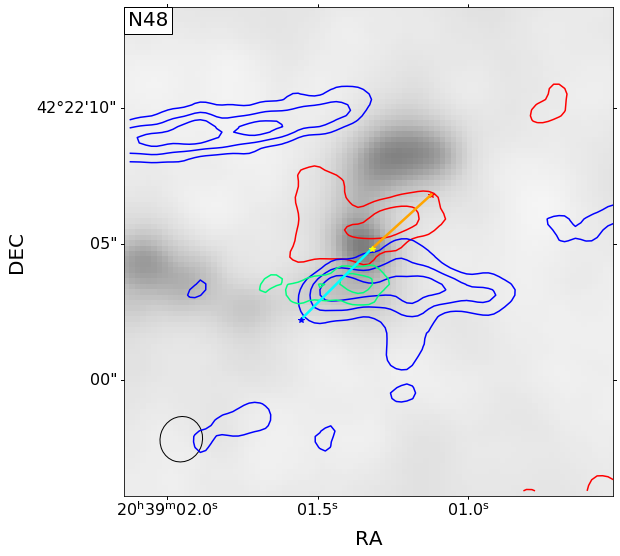}
\includegraphics[width=0.45\linewidth]{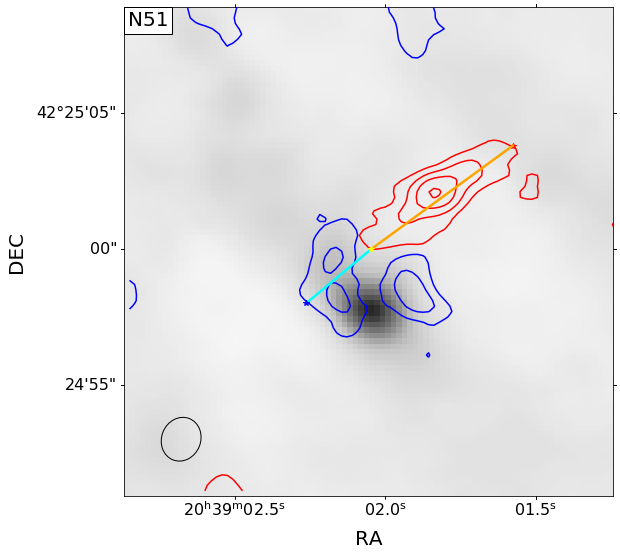}
\includegraphics[width=0.45\linewidth]{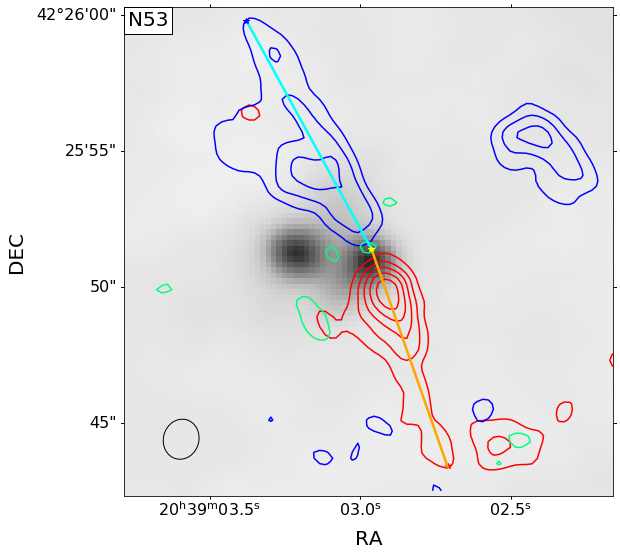}
\caption{Continuum emission of the sources of the PILS-Cygnus survey \citep{pitts2021} plotted in gray-scale ranging from --0.1 to 0.5 Jy beam$^{-1}$ in all maps. Displayed in red contours is the red-shifted CO emission, in blue contours is the blue-shifted CO emission, and in green contours is the integrated SiO emission, when detected. All contour levels are set at 3,6,9,12,15$\sigma_\text{rms}$ noise of the corresponding integrated emission.The yellow star marks the assumed origin of the outflow and the light blue and orange lines mark the extend of the outflow lobes. Finally, the beam is marked with the black contour at the bottom left.}
\label{fig:outflows2}
\end{figure*}

\begin{figure*}
\centering
\includegraphics[width=0.45\linewidth]{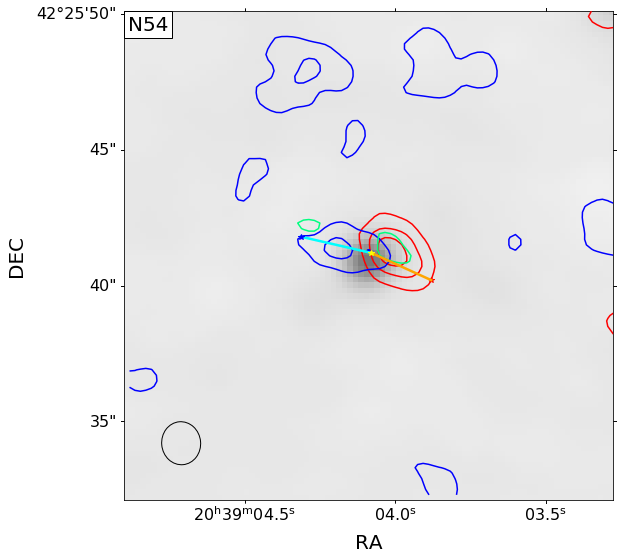}
\includegraphics[width=0.45\linewidth]{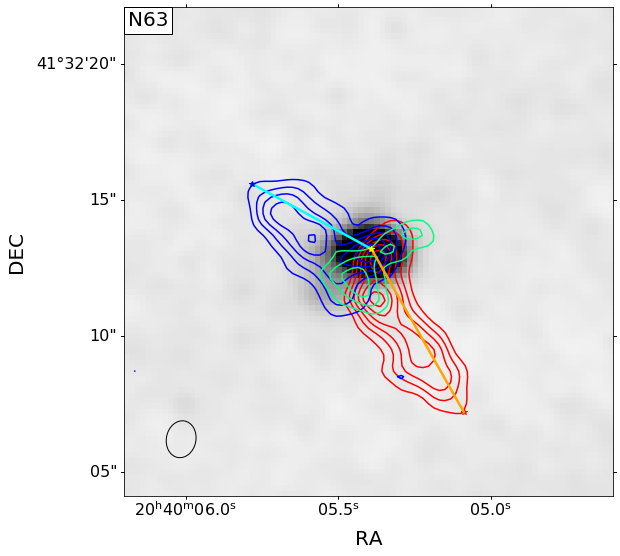}
\includegraphics[width=0.45\linewidth]{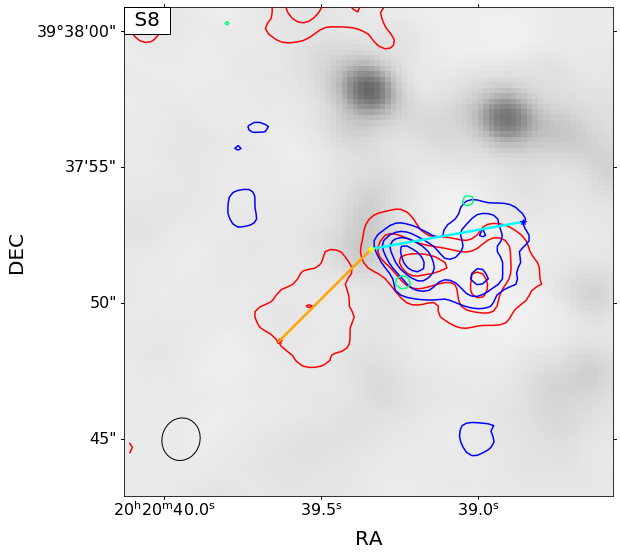}
\includegraphics[width=0.45\linewidth]{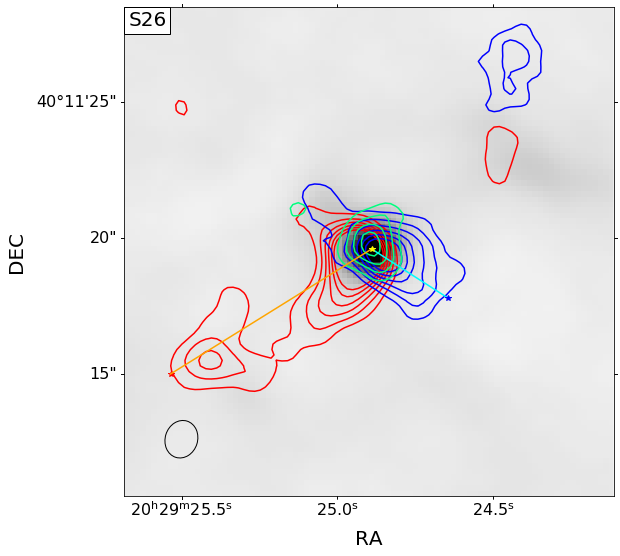}
\caption{As Fig. \ref{fig:outflows2} but for the remaining sources.}
\label{fig:outflows_no}
\end{figure*}

We detected CO 3--2 emission toward all ten sources of our sample. The line profiles show high-velocity wings extending out to $>$ 30 km s$^{-1}$ from the source velocity in the most extreme cases. Furthermore, the profiles are characterized by missing emission at velocities close to that of the source, which is likely a combined effect of optically thick emission as well as spatial filtering by the interferometer. The outflow maps are presented in Figs. \ref{fig:outflows2} and \ref{fig:outflows_no} as integrated intensity contours plotted over the corresponding continuum emission. 

The velocity ranges used for the integration of the images are given in Table \ref{table:velrange} and are relative to the source velocity. For the CO observations, we refer to the velocities closest to the source velocity as ``inner'' velocities ($v_{\text{r,in}}$ and $v_{\text{b,in}}$ for the red and blue-shifted range, respectively), while the highest velocities are called ``outer'' ($v_{\text{r,out}}$ and $v_{\text{b,out}}$). For SiO emission, a single range is used in the integration and thus only the outer limits are given, with $v_\text{r}$ being the maximum red-shifted velocity and $v_\text{b}$ the maximum blue-shifted velocity.

\begin{table*}[htbp]
\caption{Velocity limits for the integration of CO and SiO emission.} 
\label{table:velrange} 
\centering 
\begin{tabular}{l c c c c c c c} 
\hline\hline 
 \multicolumn{2}{c|}{}& \multicolumn{4}{c|}{\textbf{CO}} & \multicolumn{2}{c}{\textbf{SiO}} \\ 
\hline
Source & \begin{tabular}[c]{@{}c@{}}$v_{\text{source}}$\tablefootmark{a}\\  {[}km s$^{-1}${]}\end{tabular}& \begin{tabular}[c]{@{}c@{}}$v_{\text{r,in}}$\\  {[}km s$^{-1}${]}\end{tabular} & \begin{tabular}[c]{@{}c@{}}$v_{\text{r,out}}$ \\ {[}km s$^{-1}${]}\end{tabular} & \begin{tabular}[c]{@{}c@{}}$v_{\text{b,in}}$\\  {[}km s$^{-1}${]}\end{tabular} & \begin{tabular}[c]{@{}c@{}}$v_{\text{b,out}}$ \\ {[}km s$^{-1}${]}\end{tabular} & \begin{tabular}[c]{@{}c@{}}$v_\text{b}$\\  {[}km s$^{-1}${]}\end{tabular} & \begin{tabular}[c]{@{}c@{}}$v_\text{r}$ \\ {[}km s$^{-1}${]}\end{tabular}  \\ 
\hline
N12 &14.0& 6 & 48 & \phantom{1}--8 & --50 & --30 & 30 \\ 
N30 &11.0& 6 & 51 & --24 & --69 & --30 & 30 \\ 
N38 &--2.0& 2 & 33 & \phantom{1}--7 & --20 & --20 & 20 \\ 
N48 &--6.5& 6 & 55 & \phantom{1}--2 & --55 & --10 & 10 \\ 
N51 &--2.5& 4 & 40 & --12 & --40 & --15 & 15 \\ 
N53 &--5.5& 8 & 55 & --13 & --60 & --15 & 15 \\ 
N54 &--3.0& 3 & 53 & \phantom{1}--8 & --58 & --10 & 10 \\ 
N63 &--5.0& 15\phantom{1} & 55 & --14 & --55 & --20 & 20 \\ 
S8  &\phantom{1}1.0& 3 & 25 & --13 & --35 & --25 & 25 \\ 
S26 &--6.0& 9 & 35 & --24 & --50 & --10 & 10 \\ 
\hline 
\end{tabular}

\tablefoot{
\tablefoottext{a}{From \citet{vanderwalt2021}}}
\end{table*}

\begin{figure}
\resizebox{\hsize}{!}{\includegraphics{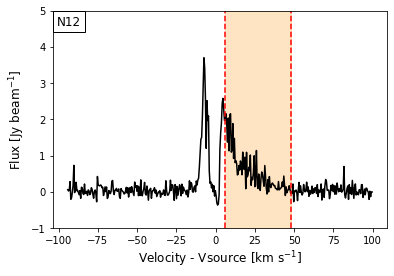}}
\caption{Spectrum of N12 from the location of the peak of the integrated red-shifted emission. The red marked area corresponds to the velocity range used in the integration of the red-shifted emission.}
\label{fig:redinteg}
\end{figure}

\begin{figure}
\resizebox{\hsize}{!}{\includegraphics{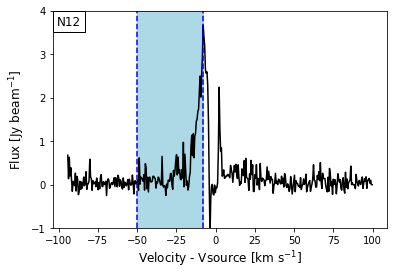}}
\caption{Spectrum of N12 from the location of the peak of the integrated blue-shifted emission. The blue marked area corresponds to the velocity range used in the integration of the blue-shifted emission.}
\label{fig:blueinteg}
\end{figure}

To estimate the velocity ranges used for the CO integrated emission maps a series of different spectra was used. First, the spectrum from the location of peak continuum emission was used for an initial velocity estimate. This velocity estimate was done as to best cover the entirety of the detected line-wings of the CO spectrum.
Next, based on the initial velocity estimates, the spectra from the location of peak blue- and red-shifted emission were extracted and used for a new velocity range estimate. This was in some cases necessary as the initial spectra did not always show clear line-wings for both blue- and red-shifted emission. 
For a final check on the velocity ranges, the spectra from the location of peak integrated blue- and red-shifted emission where used. An example of these  spectra, with the adopted velocity ranges marked in red and blue for the red- and blue-shifted part of the outflow respectively is shown in Fig. \ref{fig:redinteg} and Fig. \ref{fig:blueinteg} for N12. In addition all the spectra used for determining the velocity ranges of the ten Cygnus-X sources can be found in Appendix \ref{sec:appendixA}.

For the SiO emission, a single velocity range was used covering continuously the entire range from red- to blue-shifted emission. This velocity range was determined so as to maximize the S/N in each case and to give the best possible detection at the 3$\sigma$ level. As the SiO emission detected was overall weak and not very extended it was only used as an additional tool assisting in determining the origin of the CO outflows. 

A clear bipolar outflow structure is visible toward the majority of the sources, and in most cases this appears to be originating from the continuum sources with a few noteworthy exceptions. 
First, toward N38 only a small outflow is detected that appears to originate from a weak, secondary continuum peak, located slightly north of the primary peak. The presence of relatively strong SiO emission from the same location strengthens the argument that this is indeed the actual source of the outflow. 
N30 also appears irregular, showing a significant asymmetry between the two lobes. The red-shifted part of the outflow appears stronger and closer to the continuum peak, which we identify as the source of the outflow, while the blue-shifted part of the outflow is small and further away from the continuum peak. This we believe to be a result of missing blue-shifted emission near the source velocity, seen on the spectrum used to determine the velocity integration limits. This pushed the inner blue-shifted velocity significantly further away from the source velocity comparatively with the red-shifted one (For N30 $v_\text{b,in} = -24$ km s$^{-1}$ while $v_\text{r,in} = 6$ km s$^{-1}$).
In the case of N48, an elongated blue-shifted structure is detected slightly to the northeast of the continuum source. This structure is, as mentioned in \citep{duarte2013}, the outflow of IRS-1 and therefore not taken into account for this study.
Turning to N53, it is clear that, even though there are two distinct continuum peaks, the CO outflow is only detected from the western peak, which might suggest that the second core is not currently driving an outflow.
Finally, for the case of S8, we consider the outflow to originate from the area of extended continuum emission under the primary peak, and traveling close to the plane of the sky. In this case, one of the outflow lobes is seen in both red and blue-shifted emission near the image center, while the second lobe is only visible in the red-shifted CO emission extending toward the east of the image. In addition, a fairly strong red-shifted structure appears north of the main continuum peak, but lacks a matching blue-shifted lobe and is therefore not considered an outflow and is not used for the subsequent outflow force calculation.

\subsection{Outflow forces}

Properly calculating $F_\text{CO}$ is important for constraining the $F_\text{CO} - M_\text{env}$ and $F_\text{CO} - L_\text{bol}$ correlations. Recently \citet{vdm2013} made a comparison of the 7 most commonly used $F_\text{CO}$ calculation methods, where an inferred uncertainty of up to a factor of 6 was found. In the same work, \citet{vdm2013} suggests the use of the so-called ``separation method'' as the one least affected from observational uncertainties and it is the one used here. 

In this method, the outflow momentum ($P_\text{out}$) and the dynamical time ($t_\text{d}$) are treated as separate parameters. The dynamical time is calculated by using the maximum outflow velocity ($v_\text{max}$) and, at the same time, the momentum is calculated from the velocity-weighted integrated intensity. This approach is chosen as it offers a better estimate of the actual outflow momentum since it takes into account the kinematic structure of the outflow \citep{downes2007,curtis2010}.
The outflow force using the separation method is calculated as: 
\begin{equation}
    F_\text{CO} = c_3 \times \frac{\displaystyle K \left( \sum_\text{j} \left[ \int_{v_\text{in}}^{v_\text{out,j}}T(v')v'\text{d}v' \right]_\text{j}\right) v_\text{max}}{\displaystyle R_\text{lobe}} .
    \label{eq:eq7}
\end{equation}
Here $c_3$ is a factor that corrects for inclination effects, $K$ accounts for the mass (see Eq. \ref{eq:kappa} in the Appendix) for an assumed excitation temperature of 50 K, and the integral $\int_{v_\text{in}}^{v_\text{out,j}}T(v')v'\text{d}v'$ is the velocity-weighted integrated intensity. Finally, $R_{\rm lobe}$ is the length of the outflow lobe.
The calculation is performed for each outflow lobe individually and then summed up for the total $F_\text{CO}$.
For a detailed explanation of the calculation method and the parameters used, see Appendix \ref{sec:separationapp}. Representative results for each source, calculated for an inclination angle of 30$\degr$, are presented in Table \ref{table:separationexample} and the full results are shown in Appendix \ref{sec:outflowforcesApp} Table \ref{table:separation}.

\begin{table}
\caption{Inclination correction factors.} 
\label{table:inclinations} 
\centering 
\begin{tabular}{c c c c c c} 
\hline\hline 
$i$\tablefootmark{a} ($\degr$) & 10 & 30 & 50 & 70 & Ref \\ 
\hline
$c_1$ & 0.28 & 0.45 & 0.45 & 1.1 & 1,2 \\ 
$c_2$ & 1.6 & 3.6 & 6.3 & 14 & 1 \\ 
$c_3$\tablefootmark{b} & 0.6 & 1.3 & 2.4 & 3.8 & 3 \\ 
\hline 
\end{tabular}
\tablebib{(1) \citet{cabrit1990}; (2) \citet{cabrit1992}; (3) \citet{downes2007}}
\tablefoot{
\tablefoottext{a}{$i$ is measured from the line of sight.}
\tablefoottext{b}{Values are interpolated from Table 6 of \citet{downes2007}}, where $\alpha = 90 - i$. 
}
\end{table}

\begin{table}
\caption{Maximum outflow velocities and extents of outflow lobes.}
\label{table:maxvel} 
\centering 
\begin{tabular}{l c c c c} 
\hline\hline 
 \multicolumn{1}{c|}{}& \multicolumn{2}{c|}{Red-shifted} & \multicolumn{2}{c}{Blue-shifted} \\ 
\hline
Source & \begin{tabular}[c]{@{}c@{}}$v_{\text{max}}$\\  {[}km s$^{-1}${]}\end{tabular} & \begin{tabular}[c]{@{}c@{}}$R_{\text{lobe}}$ \\ {[}pc{]}\end{tabular} & \begin{tabular}[c]{@{}c@{}}$v_{\text{max}}$\\  {[}km s$^{-1}${]}\end{tabular} & \begin{tabular}[c]{@{}c@{}}$R_{\text{lobe}}$ \\ {[}pc{]}\end{tabular} \\ 
\hline
N12 & 28.4 & 0.060 & $-$21.5 & 0.059  \\ 
N30 & 53.3 & 0.025 & $-$51.3 & 0.054 \\ 
N38 & 10.1 & 0.018 & $-$24.3 & 0.049  \\ 
N48 & 15.1 & 0.019 & $-$27.1 & 0.023 \\ 
N51 & 13.0 & 0.041 & $-$22.9 & 0.020 \\ 
N53 & 14.9 & 0.053 & $-$39.3 & 0.060 \\ 
N54 & 30.4 & 0.015 & $-$16.6 & 0.017  \\ 
N63 & 52.4 & 0.043 & $-$45.4 & 0.032 \\ 
S8  & 17.6 & 0.030 & $-$34.2 & 0.036  \\ 
S26 & 37.0 & 0.139 & $-$40.5 & 0.053  \\ 
\hline 
\end{tabular}
\end{table}

Apart from the separation method we used two additional methods to calculate $F_\text{CO}$ for all sources in our sample. These were the $v_\text{max}$ method (Method 1) and the $\langle v \rangle$ method \citep[Method 3;][]{vdm2013}. The $v_\text{max}$ method was selected because it is the most commonly used method in the literature \citep{vdm2013} therefore presenting $F_\text{CO}$ measurements of our sample with the $v_\text{max}$ method might prove useful for further comparisons with similar works.
The $\langle v \rangle$ method, on the other hand, was selected because it is used by \citet{maud2015II} for the calculation of the outflow forces of their high-mass protostellar sample. As we aim to compare the results of our work with this sample, having the measurements in both methods will allow us to estimate if any observed difference in $F_\text{CO}$ is significant, or is inferred by the calculation method used.

The calculation of the outflow force for both methods is very similar to the separation method so we do not go into detail here. In short, for the $v_\text{max}$ method, the force is calculated as: 
\begin{equation}
        F_\text{CO} = c_1 \times \frac{\displaystyle K \left( \sum_\text{j} \left[ \int_{v_\text{in}}^{v_\text{out,j}}T(v')\text{d}v' \right]_\text{j}\right) v_\text{max}^2}{\displaystyle R_\text{lobe}} ,
    \label{eq:eq11}
\end{equation}
with $c_1$ the corresponding inclination angle correction given in Table \ref{table:inclinations} and the rest of the parameters calculated as in the separation method (see Appendix \ref{sec:separationapp}).
For the $\langle v \rangle$ method, the force is:
\begin{equation}
        F_\text{CO} = c_2 \times \frac{\displaystyle K \left( \sum_\text{j} \left[ \int_{v_\text{in}}^{v_\text{out,j}}T(v')v'\text{d}v' \right]_\text{j}\right)^2}{\displaystyle R_\text{lobe}\sum_\text{j} \left[ \int_{v_\text{in}}^{v_\text{out,j}}T(v')\text{d}v' \right]_\text{j}} ,
    \label{eq:eq12}
\end{equation}
with $c_2$ again being the corresponding inclination correction factor, given in Table \ref{table:inclinations}. The rest of the calculation is performed as for the separation method, with the only difference that $v_\text{max}$ is not needed in this case. The outflow forces, as measured with the $v_\text{max}$ and $\langle v \rangle$ methods, are presented in Appendix \ref{sec:outflowforcesApp} (Tables \ref{table:vmax} and \ref{table:vmean}, respectively).

\begin{table}
\caption{$F_\text{CO}$ calculated with the separation method.} 
\label{table:separationexample} 
\centering 
\begin{tabular}{l c c c} 
\hline\hline 

Source & $F_\text{red}$ & $F_\text{blue}$ & $F_\text{total}$ \\
\hline
N12 & \phantom{1}2.4 & \phantom{1}4.8 & \phantom{1}7.2 \\
N30& 159.8\phantom{1} & 29.3 & 189.1\phantom{1} \\ 
N38& \phantom{1}0.6 & \phantom{1}2.0 & \phantom{1}2.6 \\ 
N48& \phantom{1}3.0 & \phantom{1}5.9 & 8.9 \\ 
N51& \phantom{1}1.0 & \phantom{1}6.1 & 7.1 \\ 
N53& \phantom{1}0.7 & 16.1 & 16.8 \\ 
N54& \phantom{1}3.8 & \phantom{1}1.9 & \phantom{1}5.7 \\ 
N63& 17.9 & 25.3 & 43.2 \\ 
S8& \phantom{1}4.9 & 10.4 & 15.3 \\ 
S26& 32.5 & 58.9 & 91.4 \\ 
\hline 
\end{tabular}
\tablefoot{The forces are calculated for an inclination angle of $i=30$\degr\ and they are in units of 10$^{-5}$ M$_\odot$ yr$^{-1}$ km s$^{-1}$ .}
\end{table}

\begin{figure}[h]
\centering
\resizebox{\hsize}{!}{\includegraphics{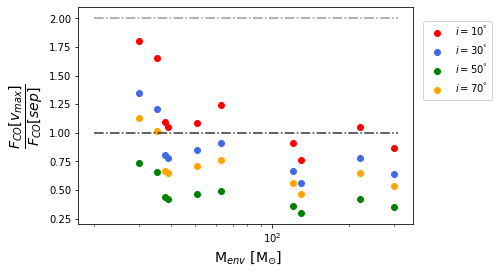}}
\resizebox{\hsize}{!}{\includegraphics{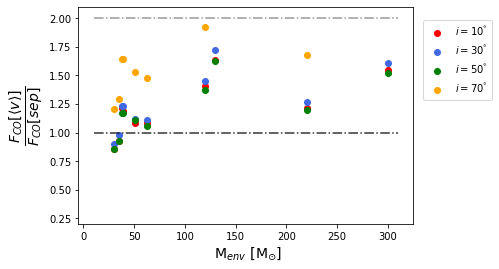}}
\caption{Comparison of different outflow force calculation methods. [Top.] Ratio of the outflow forces for the 10 sources of the PILS-Cygnus survey, as measured using the $v_\text{max}$ over the separation method. [Bottom.] Ratio of the outflow forces for the 10 sources of the PILS-Cygnus survey, as measured using the $\langle v \rangle$ over the separation method. In both cases, red points mark the results for $i=10\degr$, blue for $i=30\degr$, green for $i = 50\degr$ and orange for $i=70\degr$. Also plotted in black and gray dotted lines are the 1:1 and 2:1 ratios respectively.} 
\label{fig:vmaxcomparison}
\end{figure}

In Fig. \ref{fig:vmaxcomparison} the ratio of outflow forces as measured using the different calculation methods for the 10 sources of the PILS-Cygnus survey are plotted as a function of their envelope masses. This comparison reveals small differences between the different methods, with the results varying by a factor of $<$3. This result falls well within the factor of 6 that \citet{vdm2013} reports among all calculation methods discussed therein. 

Focusing on the comparison between the separation method and the $\langle v \rangle$ method reveals a difference between the methods that is up to a factor of $\sim$2. The very small difference found between the two methods is important as it means that a direct comparison between results from \citet{maud2015II} and other works, using the separation method, is possible and will not be significantly affected by the choice of outflow force calculation method.

Regarding the different inclination angles, for all three methods, we find that the difference, even between the most extreme cases of going from $i = 10\degr$ to $i = 70\degr$, is smaller than an order of magnitude. Therefore, even though constraining the precise inclination angle of an outflow is difficult, this uncertainty does not affect the measured outflow force significantly.

\subsubsection{Uncertainties on the outflow forces}

It is important to highlight that the outflow force measurements for the PILS-Cygnus sources likely are lower limits of the true $F_\text{CO}$. This is a result of the lower velocity limits used for the integration in Eq. \ref{eq:eq7} (the inner velocities in Table \ref{table:velrange}). Due to the optical thickness of the CO emission in combination with spatial filtering effects arising from the nature of interferometric observations, a velocity range of $\approx 10$  up to $\approx 30$ km s$^{-1}$ surrounding  $v_\text{source}$ is excluded from the calculation. Therefore, the low-velocity outflowing gas, that holds a significant part of the outflow mass \citep{dunham2014}, is not taken into account in measuring $F_\text{CO}$.
The outflow force scales with the square of the outflow velocity $v^2$, but depends only linearly on the outflow mass. As a result, the higher-velocity material contributes significantly more to the total outflow force compared to the slow moving material. Thus, not taking into account the smaller velocities of the outflow in the calculation of the outflow force likely has a smaller impact on the resulting outflow forces. 
More precisely, \citet{offner2011} found that integrating over velocities $> 2$ km s$^{-1}$ leads to an underestimate of the outflow mass by a factor of 5 - 10 from simulations, a result that was confirmed observationally later by \citet{dunham2014}.
Interestingly, N30, S26, and N63, the sources driving the strongest outflows, are also the sources with the largest masked-out velocity range. This is most likely due to, based on the $F_\text{CO}$ - $M_\text{env}$ correlation, these sources having larger envelopes and therefore being susceptible to spatial filtering effects at higher velocities, which in turn forces the inner velocities limits to higher velocities (in absolute for the blue-shifted ones).

We do not expect the difference in $F_\text{CO}$ to be significant, but even if our measurements underestimate the true $F_\text{CO}$  by as much as an order of magnitude, the results for the PILS-Cygnus sources would still fall within the scatter of the high-mass sample from \citet{maud2015II}. Therefore, although the outflow measurements presented in Table \ref{table:separation} are lower limits of the true $F_\text{CO}$, our results regarding clustered star formation remain unchanged.

\subsubsection{Sample biases}
\label{sec:biases}
An important issue that needs to be addressed when studying samples such as here, are the selection criteria used for the observations of the high- and low-mass protostellar samples (Fig. \ref{fig:forcemasscorel}  and Fig. \ref{fig:forcelumcorel}). The sample of \citet{yildiz2015} was specifically chosen to have strong and bright outflows \citep[the original sample was from a \emph{Herschel} key program; see ][ for more details]{vandishoeck2011}. Similarly, the sample from \citet{mottram2017} was based on a luminosity cut-off to target the brightest protostars in nearby clouds, which presumable coincides with the sources driving the strongest outflows. As a result, these samples are most likely representative of the upper limit of the actual low-mass distribution. In contrast, the sample presented in \citet{maud2015II} has no such biases. Including a more representative sample of low-mass sources might allow for a better connection of the high- and low-mass samples, which currently appear separate. In addition, such a sample would of course yield more accurate statistics for the correlations followed by the protostellar samples.

\section{Discussion}
\label{sec:discussion}

\subsection{Outflow force vs $M_\text{env}$ and $L_\text{bol}$}
\label{sec:forcelumcorel}

To better be put into context, the resulting outflow forces are compared with the correlations of the outflow force with the envelope mass and bolometric luminosity already established for protostellar sources \citep[e.g.,][]{bontemps1996, beuther2002}. Specifically, this comparison is first performed for the $F_{\rm CO}$ -- envelope mass correlation for protostars and then it is expanded to include extragalactic sources when comparing to the luminosity.

\subsubsection{$M_\text{env}$ determination}
\label{sec:forcemasscorel}

As demonstrated by \citet{bontemps1996}, outflow forces correlate with the mass of the protostellar envelope. The envelope masses used for the Cygnus-X sources, shown in Table \ref{table:cygnusmasslum}, are mostly taken from \citet{pitts2021} who used SED fitting of \emph{Herschel} observations assuming a distance of 1.4 kpc, the average distance to the Cygnus-X sources \citep{rygl2012}, with the exception of S26 that was initially calculated from \citet{pitts2021} for the distance of 3.3 kpc \citep{rygl2012}, also assumed in this work. For sources N38 and N54, no envelope masses are given in \citet{pitts2021}, therefore previous measurements from \citet{motte2007} are used. These are scaled to the assumed distance of 1.4 kpc.

\citet{pitts2021} used SCUBA 450 $\mu$m observations to determine the size of the envelopes in order to measure their mass. These observations have a resolution of 14 arcesonds, significantly larger than the $< 2$ arcseconds of the SMA observations used to map the CO outflows. These SMA observations reveal multiple substructures, that are included in the same envelope, but are not all driving outflows. In such cases, the entire envelope mass is attributed to the single observed outflow source even though it is likely to break down into multiple sources driving separate outflows. A clear example of this is N53, where the continuum emission (Fig. \ref{fig:outflows2}) shows two separate cores, but the CO contours reveal that only one of them actually drives an outflow and is therefore assigned the entire envelope mass.
Similar cases are the outflows of S8, N48 and N38.
Especially for N38, the origin of its outflow is located at a secondary continuum peak north of the primary continuum peak. This structure remains unresolved in the observations used by \citet{motte2007} to estimate the envelope mass of the source. It is therefore reasonable to assume that the envelope mass found in \citet{motte2007} is mostly representative of the primary continuum peak, and not of the actual outflow source.
In an attempt to use a more representative estimate of the actual envelope mass of the source behind the detected outflow we scaled down the mass found by \citet{motte2007} by a factor of 4, and used this new value (see Table \ref{table:cygnusmasslum}) for this work. The scaling factor of 4 was determined by the ratio of the intensity of continuum emission of the primary and secondary peak. We recognize of course that this estimate is very uncertain, and that the actual envelope mass of N38 as a result is highly uncertain.

\subsubsection{Correlation with envelope mass}
\begin{figure*}
    \centering
    \includegraphics[width=17cm]{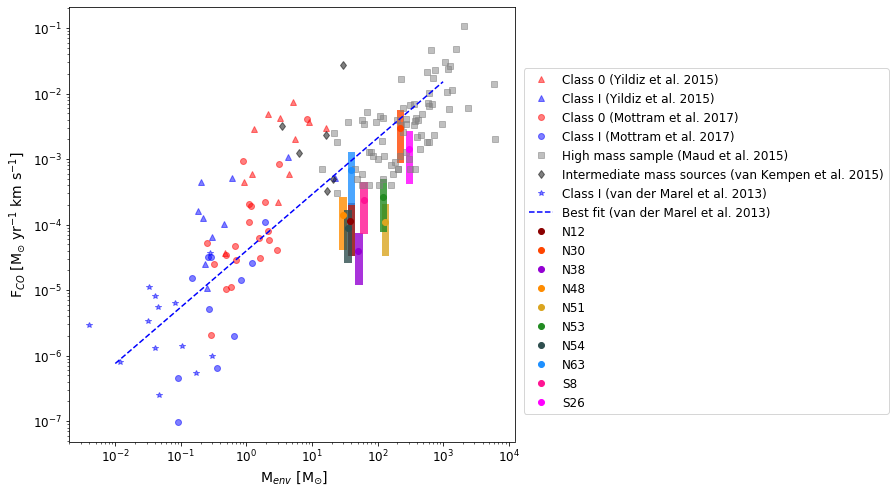}
    \caption{ Outflow forces for the ten sources of the PILS Cygnus survey over their envelope mass. The outflow forces are plotted using colored rectangles covering the range of possible results for the different inclination correction factors. Plotted in the same figure are, Class 0 sources from \citet{yildiz2015} in red triangles, Class I sources from \citet{yildiz2015} in blue triangles, Class 0 sources from \citet{mottram2017} in red circles, Class I sources from \citet{mottram2017} in blue circles, Class I sources from \citet{vdm2013} in blue stars, intermediate-mass sources from \citet{vankempen2009} in dark gray diamonds and high mass sources from \citet{maud2015II} in light gray squares. The blue dashed line shows the best fit as found in \citet{vdm2013} for CO $J$ = 3-2 observations of a low mass protostellar sample.
    }
    \label{fig:forcemasscorel}
\end{figure*}

To explore the $F_\text{CO}$ - $M_\text{env}$ correlation we plot in Fig. \ref{fig:forcemasscorel} the outflow forces versus the envelope mass for the ten sources in the Cygnus-X molecular cloud. 
These represent the extent of the resulting outflow forces for the four different inclination correction factors. Alongside our results, the low-mass protostellar sample is plotted. 
Finally, the high-mass sample from \citet{maud2015II} is plotted. 
All four of these studies use CO $J$ = 3--2 observations, similar to this study.  
We also note that \citet{yildiz2015}, \citet{mottram2017} and \citet{vdm2013} use the separation method for calculating $F_\text{CO}$ but \citet{maud2015II} uses the $\langle v \rangle$ method. As previously discussed, the $\langle v \rangle$ method, on average, varies from the separation method by up to a factor of 2. This uncertainty is significantly smaller than the inherent uncertainties in the $F_\text{CO}$ measurements therefore a direct comparison of the different measurements is reasonable.
Finally, the blue dashed line represents the best-fit on the correlation as calculated by \citet{vdm2013} for a sample of low-mass protostars.

The PILS-Cygnus survey sources are, in their majority, tightly packed in the same molecular cloud structure and located in close proximity to the Cyg-OB2 association, one of the largest OB associations in our Galaxy; they therefore are a prime example of clustered star formation. On the other hand, \citet{maud2015II} specifically targeted more isolated high-mass sources.
The good agreement between the results of the two studies for sources of similar envelope masses, as can be seen in Fig. \ref{fig:forcemasscorel}, suggests that clustered star formation does not significantly impact the outflow activity of protostars. 
An exception to this may be N38, which appears to have a slightly weaker outflow than sources of similar mass. As discussed in Sec. \ref{sec:maps}, N38 has only a compact outflow, that originates from a secondary continuum peak close to the main peak which causes great uncertainty for its true envelope mass. Even so, given the large scatter of the high-mass sample, N38 largely still follows the same correlation as the rest of the sample.

It is believed that the outflow force is directly connected to the mass accretion rate of the driving source \citep[e.g.,][]{bontemps1996,duarte2013}. 
Therefore, the previous result, that clustered star formation does not significantly affect protostellar outflow activity, suggests that the accretion process itself is not affected by the surrounding environment of the protostar but is rather dominated by local processes. 
Of course, the sample of clustered sources examined is very limited, containing only 10 sources all taken from the same star-forming region. In order to verify this result further studies of clustered protostellar sources are required.

Additionally, in Fig. \ref{fig:forcemasscorel}, a clear distinction can be seen between the low- and high-mass protostellar sources, highlighted by the lack of measurements for sources with $M_\text{env}$ $\approx$ 10 M$_\odot$. The few intermediate-mass sources plotted from \citet{vankempen2009} seem more in line with the low-mass sample but the sample consists of too few sources to allow for any concrete conclusions. Properly connecting the two samples will assist significantly in determining if they actually follow the same correlation or not.

\subsubsection{Correlation with luminosity}
\label{sec:lbolcorrel}

\begin{figure*}
    \centering
    \includegraphics[width=17cm]{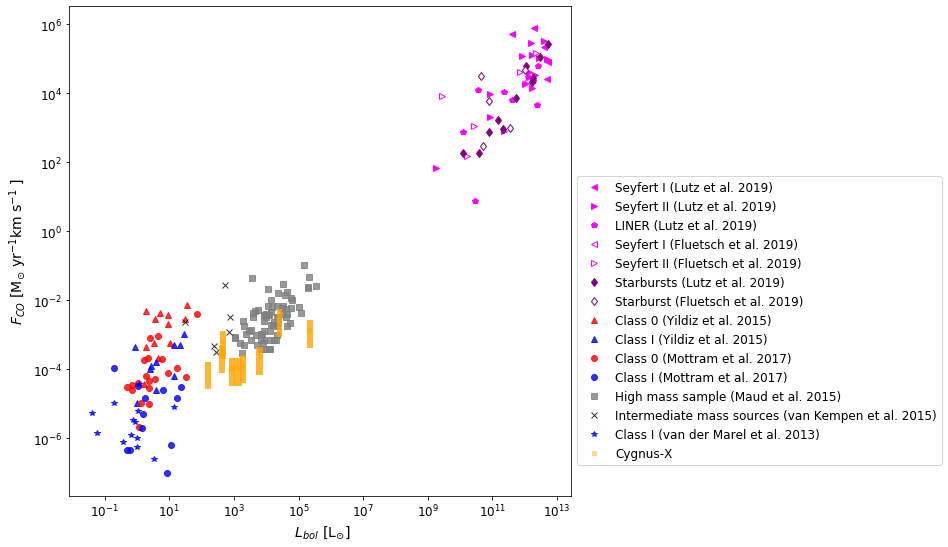}
    \caption{Outflow forces over bolometric luminosities for a sample of low-mass protostars, a sample of high-mass protostars, and extragalactic sources. Red points mark Class 0 low-mass protostars, with triangles taken from \citet{yildiz2015} and circles from \citet{mottram2017}. Blue points mark Class I low-mass protostars, again with triangles from \citet{yildiz2015}, circles from \citet{mottram2017}, and stars from \citet{vdm2013}. gray squares mark the high-mass protostellar sample from \citet{maud2015II} while gray crosses mark the intermediate mass sources described in \citet{vankempen2009}. The Cygnus-X sources, measured in this work, are shown with orange rectangles covering the extent of their outflow forces for the different available inclination correction factors. For the extragalactic sample, AGN sources are marked in magenta, with left-looking triangles for Seyfert I, right-looking triangles for Seyfert II and pentagons for LINER sources. Purple diamonds mark the starburst sources. Finally, filled points represent measurements taken from \citet{lutz2020} and empty ones from \citet{fluetsch2019}.}
    \label{fig:forcelumcorel}
\end{figure*}

To explore the $F_\text{CO}$ - $L_\text{bol}$ correlation we used a sample of protostellar and extragalactic sources collected from the literature in which we added the outflow forces measured in this work for the Cygnus-X sources. 
The protostellar sample consists of low-mass protostellar sources from \citet{vdm2013}, \citet{yildiz2015} and \citet{mottram2017} and high-mass protostars from \citet{maud2015II} and this work, while the extragalactic sample is taken from \citet{lutz2020} with additional sources taken from \citet{fluetsch2019}. This sample includes both AGN galaxies, split into Seyfert I, II, and LINERs, and starburst (SB) galaxies.
Both \citet{lutz2020} and \citet{fluetsch2019} used the time-averaged thin-shell approach for calculating the outflow force. In this approach, a constant outflow rate over the entire life-time of the outflow ($\displaystyle t_\text{flow} = R_\text{out}/v_\text{out}$) is assumed, yielding a final density profile of $\displaystyle \rho \propto r^{-2}$.
The outflow rate is then given as: 
\begin{equation}
    \dot{M}_\text{out} = \frac{M_\text{out} v_\text{out}}{R_\text{out}} ,
\end{equation}
and the outflow force is calculated as: 
\begin{equation}
    F_\text{CO} = \dot{M}_\text{out} v_\text{out} .
\end{equation}
This approach is very similar to the $v_\text{max}$ method \citep{vdm2013} often used to calculate protostellar outflows where a unique velocity is assumed for the entirety of the outflow.
Figure \ref{fig:forcelumcorel} shows the outflow forces over the corresponding bolometric luminosities for all sources.
The $L_\text{bol}$ values used for the Cygnus-X sample, presented in Table \ref{table:cygnusmasslum}, are mostly from \citet{pitts2021}, with the exception of N38 and N54 which are taken from \citet{Irene2015} and \citet{kryukova2014} respectively.

In Fig. \ref{fig:forcelumcorel} the same characteristics displayed in Fig. \ref{fig:forcemasscorel} (i.e., good agreement between the Cygnus-X and the \citet{maud2015II} sources and a clear separation between the low- and high-mass sample) can be seen. We note though that the separation between the low- and high-mass samples appears more pronounced in the $F_\text{CO}$ - $L_\text{bol}$ correlation (Fig. \ref{fig:forcelumcorel})  than in the $F_\text{CO}$ - $M_\text{env}$ correlation (Fig. \ref{fig:forcemasscorel}). Therefore, overall this comparison reinforces the points made in Sec. \ref{sec:forcemasscorel}.

The most important questions arising from Fig. \ref{fig:forcelumcorel} is how well  the $F_\text{CO}$ - $L_\text{bol}$ correlations for the different object types compare, and if it is at all possible to assume a common correlation for all three (low-mass, high-mass and extragalactic) samples.
In order to address these questions, we calculated the best-fit of the $F_\text{CO}$ - $L_\text{bol}$ correlation for each of the samples. 
In addition to these, we also examined the AGN- and SB-driven outflows separately, as well as the Class 0 and Class I sources.
The least-squares method is used to find the best-fits, which are of the form: 
\begin{equation}
    \log(F_\text{CO}) = a \log(L_\text{bol}) + b .
    \label{eq:bestfit}
\end{equation}
The $a$ and $b$ parameters describing the fits for the various subsamples are shown in Table \ref{table:bestfits} and the correlations themselves are shown in Fig. \ref{fig:fits} in the Appendix.

\begin{table}[htbp]
\caption{Best-fit parameters for the various subsamples.} 
\label{table:bestfits} 
\centering 
\begin{tabular}{l c c } 
\hline\hline 
Sources & Slope ($a$) & Intercept ($b$) \\ 
\hline
Extragalactic  & $0.93 \pm 0.11$ & $-6.7 \pm 1.3$ \\ 
AGN  & $0.86 \pm 0.14$ & $-5.9 \pm 1.6$ \\ 
SB  & $1.08 \pm 0.20$ & $-8.6 \pm 2.3$ \\
High-mass & $0.69 \pm 0.08$ & $-5.5 \pm 0.3$ \\
Low-mass & $0.89 \pm 0.20$ & $-4.7 \pm 0.2$ \\
Class I & $0.52 \pm 0.24$ & $-5.1 \pm 0.2$ \\
Class 0 & $1.02 \pm 0.26$ & $-4.3 \pm 0.2$ \\
\hline 
\end{tabular}
\tablefoot{The best fit is of the form $\log(F_\text{CO}) = a \log(L_\text{bol}) + b$}
\end{table}

To properly compare the various fits a statistical approach is needed. 
As a first step, we performed an $F$-test on the variances of the samples, comparing them in pairs.
Briefly, to perform an $F$-test we need to calculate the $F$-value, which is given as:
\begin{equation}
    F = \frac{s_1}{s_2} ,
\end{equation}
where $s_1$ and $s_2$ are the variances to be compared, with $s_1 > s_2$. The $F$-value is then compared with a critical value ($F_\text{crit}$), which depends on the degrees of freedom (DF) of both samples. These are calculated as \mbox{$N -1 $} for each sample, with $N$ being the number of sources in the sample. If the $F$-value calculated is smaller than $F_\text{crit}$, then the two variances are consistent with being similar and the two samples can be compared. In this work we use the $F$-table for $\alpha = 0.05$, which corresponds to a 95\% confidence level.

For the pairs with comparable variances we then performed two-tailed $t$-tests in order to examine if the parameters of their best-fit, the slope ($a$) and the intercept ($b$), are similar or not. For a $t$-test the initial hypothesis is that the parameters are equal (in this case the intercept, $H_0$: $b_1 = b_2$) and we calculate the t-value as: \begin{equation}
    t = \frac{b_1 - b_2}{\sqrt{s_{b_1} + s_{b_2}}} ,
\end{equation}
where $b_1$ and $b_2$ are the compared parameters of the two fits, and $s_{b_1}$ and $s_{b_2}$ the corresponding variances. The $t$-value is then used alongside the degrees of freedom, which is in this case is $N_1 + N_2 - 4$ with $N_1$ and $N_2$ the sizes of the two samples whose fits we are comparing, in order to calculate the $P$-value.

If the resulting $P$-value is lower than the confidence level $\alpha$ then the initial hypothesis can be rejected and the compared parameters are not equal. On the other hand, if the $P$-value exceeds $\alpha$ then the initial hypothesis cannot be rejected and we assume that the two parameters are equal. The confidence level is again $\alpha = 0.05$ resulting in a 95\% level of confidence in our results.

We present in Table \ref{table:Fvalues} the $F$-values calculated for all comparisons among the seven samples as well as the degrees of freedom corresponding to each sample.
Comparing these values with the appropriate critical values from the $F$-table we were able to determine which samples are comparable with each other and are marked in boldface in Table \ref{table:Fvalues}. Some of the most interesting results though are that the variances for the three extragalactic samples (the entire sample, AGN, and SB) are comparable with one another, that the high-mass protostellar sample variance is only comparable with the SB sample, and also that the Class 0 sample is comparable with all three extragalactic samples.

\begin{table*}[]
\caption{$F$-values for the best-fit variances comparisons and the degrees of freedom for each sample.} 
\label{table:Fvalues}
\centering 
\begin{tabular}{l| c c c c c c c} 
\hline\hline 
 & Class 0 & Class I & Low-mass & High-mass & Starburst & AGN  & DF\tablefootmark{a}\\ 
\hline
Extragalactic &\textbf{1.11}&1.85&2.11&2.28&\textbf{1.51}&\textbf{1.09}& 49 \\
AGN & \textbf{1.01} & \textbf{1.69} & 1.93 & 2.49 & \textbf{1.66} & - & 33 \\ 
Starburst & \textbf{1.67} & 2.81 & 3.19 & \textbf{1.50} & - & - & 15 \\
High-mass & 2.52 & 4.22 & 4.80 & - & - & - & 84 \\ 
Low-mass & 1.91 & \textbf{1.14} & - & - & - & - & 66\\ 
Class I & \textbf{1.67} & - & - & - & - & -& 34 \\ 
Class 0 & - & - & - & - & - & -& 31 \\ 
 \hline
\end{tabular}
\tablefoot{ 
Mirrored comparisons are equivalent and therefore not shown. Values in boldface are statistically significant meaning that the two populations are comparable. \tablefoottext{a}{DF = $N - 1$ for the $F$-test}
}

\end{table*}

\begin{table}[]
\caption{$P$-values and DF for comparison of the intercept and slope of samples with matching variances.} 
\label{table:tvalues} 
\centering 
\begin{tabular}{l| c c c} 
\hline\hline 
 & $P_\text{slope}$ & $P_\text{inter}$ & DF\tablefootmark{a} \\
\hline
\textbf{Extragalactic - AGN} &0.72 & 0.69\phantom{1} & 80 \\ 
\textbf{Extragalactic - SB}  & 0.53 & 0.49\phantom{1} & 62 \\ 
\textbf{Extragalactic - Class 0} & 0.75 & 0.07\phantom{1} & 78 \\
\textbf{AGN - Starburst} & 0.40 & 0.35\phantom{1} & 46 \\
\textbf{AGN - Class I} & 0.21 & 0.65\phantom{1} & 65 \\
\textbf{AGN - Class 0} & 0.61 & 0.33\phantom{1} & 62 \\ 
\textbf{Starburst - Class} 0 & 0.87 & 0.07\phantom{1} & 44  \\ 
\textbf{Starburst - High-mass} & 0.08 & 0.19\phantom{1} & 97 \\ 
\textbf{Low-mass - Class I} & 0.23 & 0.10\phantom{1} & 98 \\
Class I - Class 0 & 0.15 & 0.003& 63\\
 \hline 
\end{tabular}
\tablefoot{
\tablefoottext{a}{DF = $N_1 + N_2 - 4$ for the two-tailed $t$-test}. Pairs marked in boldface can be considered to follow the same $F_\text{CO}$ - $L_\text{bol}$ correlation}
\end{table}

We performed $t$-tests for the best-fits of the sample pairs found having comparable variances and present the resulting $P$-values along with the corresponding DF in Table \ref{table:tvalues}. The $P_\text{slope}$ value corresponds to the comparison of the slopes of the fits, while $P_\text{inter}$ to the comparison of the intercepts. All pairs marked in boldface in Table \ref{table:tvalues} can statistically be considered to follow the same correlation.

At this point it is also important to note that two methods are often used for calculating outflow forces from extragalactic sources. In the first, the outflow force is calculated assuming a time-averaged thin shell approximation \citep[e.g.,][]{lutz2020, fluetsch2019} (described earlier in Sec. \ref{sec:lbolcorrel}). In the second method, the outflowing gas is assumed to fill the entire cavity of the outflow with a constant density \citep[e.g.,][]{cicone2014,fiore2017}. This alternate calculation method is found to differ by a factor of three compared to the thin shell approach \citep{lutz2020}.
Both extragalactioc samples used in this work use the thin shell method, therefore in order to test if the selected method of outflow-force calculation has any impact on our results, we increased the outflow forces of the extragalactic sample by a factor of three, and performed again the statistical analysis for the new best-fits.
The resulting differences in $F$-values where minimal ($< 10^{-4}$) but the $P$-values concerning the comparison of the estimated intersections varied by up to 30\% in some cases. Even the largest of these changes did not have a statistically significant effect on any of our results. 
Therefore, at this point, there is no apparent preference toward either of the methods used for calculating outflow forces, and the following conclusions appear to stand for both scenarios.

The first interesting result of this statistical comparison is that, while \citet{maud2015II} argue that outflow forces from low-mass Class I sources and high-mass sources follow the same correlation with bolometric luminosity, based on an extrapolation of the two correlations, here it is found that the two samples, Class I and high-mass outflows, show significantly different variances, and thus they are likely drawn from two different populations. Going beyond the Class I outflows, the variance is also significantly different between the Class 0 and total low-mass outflows when compared to the high-mass outflows. Specifically, the low-mass outflows show larger variances compared to their high-mass counterparts. This discrepancy is unlikely to come from the use of different calculation methods for the outflow force, as this only adds an offset in a log-log plot. Instead, it is possible that a significant mass fraction of the high-mass outflows is CO-dark, which means that it is not traced by CO lines. This could be in the form of C$^+$ \citep{leurini2015,gusdorf2015}, where the ionization is driven by the higher UV-luminosity of high-mass sources. An alternative is that the low-mass outflows inherently show more variation, for example due to episodic accretion \citep[e.g.,][]{audard2014}. However, exploring these scenarios in detail is beyond the scope of this work.

The lack of agreement between the low- and high- mass samples found here is in agreement with the apparent separation of the two samples detected in both Figs. \ref{fig:forcemasscorel} and \ref{fig:forcelumcorel} and discussed in Sec. \ref{sec:forcemasscorel}.

The most interesting result arising from this comparison though is that all the different extragalactic sources appear to follow the same correlation as low-mass Class 0 protostellar sources. A result that hints toward a connection between the two otherwise very different samples.

Additionally, an intriguing question that comes from the apparent agreement between high-mass protostellar outflows and extragalactic outflows from starburst galaxies, found in the previous statistical comparison
, is whether this is due to the dominant role of high-mass protostellar outflows in the total extragalactic outflow activity. 
To address this, an order-of-magnitude calculation was done, assuming an initial mass function (IMF) of \citep{kroupa2002}: 
\begin{equation}
    \xi(M_*) = 
    \begin{cases}
    (M_*/0.08 M_\odot)^{-0.3},~~ 0.01 \le M_* / M_\odot < 0.08 \\
    (M_*/0.08 M_\odot)^{-1.3},~~ 0.08 \le M_* / M_\odot < 0.5 \\
    0.09(M_*/0.5 M_\odot)^{-2.3},~~ 0.5 \le M_* / M_\odot < 1.0 \\
    0.02(M_*/0.5 M_\odot)^{-2.7},~~ 1.0 \le M_* / M_\odot \\
    \end{cases} ,
\end{equation}
where $\xi$ is the number of stars for a given stellar mass $M_*$.
The IMF is defined for main-sequence stellar masses whereas the masses shown in Fig. \ref{fig:forcemasscorel} are envelope masses. In order to translate stellar masses to envelope masses a core-efficiency of $\sim$ 30\% is assumed \citep[e.g.,][]{Andre2010}, that is, approximately 30\% of the envelope or core mass ends up in the star. While this assumption comes with a number of caveats \citep[see, e.g.,][for a full discussion]{offner2014}, it is the best available at this moment.
To get the contribution to the overall outflow force from protostellar sources with $M_\text{env}$ we weighted the outflow force with the number of stars at a given envelope mass, where the outflow force is estimated from the best-fit presented in \citet{vdm2013}: 
\begin{equation}
    \log(F_\text{CO}) = -4.4 + 0.86\log(M_\text{env}) \ .
\end{equation}
We selected this fit due to the good agreement it shows with the high-mass sample, as can be seen in Fig. \ref{fig:forcemasscorel}.
To get the total outflow contribution for the two samples, the result is integrated over $0.03 \leq M_\text{env}/M_\odot < 30$ for the low-mass sample and over $30 \leq M_\text{env}/M_\odot < 3 \times 10^{4}$ for the high-mass sample.
The resulting outflow forces show that low-mass protostellar sources are responsible for $\sim 95\% $ of the total protostellar outflow force, while high-mass sources are responsible for the remaining $\sim$ 5\%.

The above result is based on a comparison between outflow forces and envelope masses, but a similar correlation exist between the outflow force and the luminosity. The question then is whether this correlation will lead to the same conclusion. \citet{pitts2021} compared envelope masses and bolometric luminosities for protostars with luminosities in the range of $<$ 1 $L_\odot$ to $>$ 10$^5$ $L_\odot$, and found that they correlate with envelope mass as $L_\text{bol} \propto M_\text{env}^{1.2}$. Thus, since the power-law index is shallower than that of the IMF at $M$ $>$ 0.5 $M_\odot$, we would also conclude that the low-mass protostars drive the outflow contribution.

The correlations are based on observations of a number of high-mass star-forming regions, which likely contain both high- and low-mass protostars. In these cases, the regions contain a single dominant outflow that is driven by the high-mass protostar, a result that is common to other studies \citep[e.g.,][]{beuther2002}. This would suggests that the outflows from high-mass protostars dominate the output from clusters, in contradiction with the result above. However, the time over which high-mass sources accrete and drive outflows is significantly shorter than that of their low-mass counterparts \citep[$\sim$ 10$^5$ vs 5$\times$10$^5$ years; e.g.,][and references therein]{motte2018, dunham2014}. Furthermore, it is unclear when low-mass stars form in clusters compared to the high-mass stars, more precisely whether they form at the same time or if the low-mass stars form before or after the high-mass stars. In this analysis, we are looking at the integrated effect of the outflows from all stars formed in a cluster, implying that at certain points in time the outflow from a single high-mass protostar may very well dominate the outflow output.

Even so, this result, although an order-of magnitude estimate, suggests that it might be low-mass protostars that dominate the total extragalactic outflow activity especially in the case of the starburst galaxies, where the outflows are believed to originate from forming stars. Therefore, the detected agreement of starburst galaxies and high-mass protostellar sources cannot be easily attributed to the dominance of high-mass protostellar outflows when averaging over a large number of clusters in the total outflow activity of starburst galaxies, unless the initial mass function is significantly more top-heavy in these galaxies.

\section{Summary and conclusions}
\label{sec:conclusion}
This paper presents interferometric observations of CO $J$ = 3--2 and SiO $J$ = 8--7 emission lines of ten high-mass protostellar sources in the Cygnus-X molecular cloud, taken from the PILS-Cygnus survey and carried out with the SubMillimeter Array.
In addition, an extended sample of low-mass protostellar, high-mass protostellar and extragalactic outflows with measured outflow forces, was gathered from the literature. 
Using these data, the main conclusions arising from this work, are as follows:
\begin{itemize}
    \item Using integrated intensity maps of the CO $J$ = 3--2 line emission, molecular outflows were detected originating from all ten sources of the PILS-Cygnus survey, and their respective outflow forces were successfully measured using both the separation, the $\langle v \rangle$, and the $v_\text{max}$ methods.
    \item The close agreement found between the $F_\text{CO}$, measured for the ten Cygnus-X sources, with results from the literature for sources with similar $M_\text{env}$ and $L_\text{bol}$, indicate that clustered star formation has no significant effect on the outflow behavior of protostars. As the outflow activity is tied to the accretion process, this result suggests that protostellar accretion is not significantly affected by the surrounding environment, but it is rather tied to local processes, determined by the protostar itself.
    \item Examination of the $F_\text{CO}$ - $M_\text{env}$ and $F_\text{CO}$ - $L_\text{bol}$ correlations for the protostellar sources, reveals clear differences between the low- and high-mass protostellar sources both in the correlation itself as well as in the observed scatter.
    \item The statistical comparison of the best-fits to the $F_\text{CO}$ - $L_\text{bol}$ correlation, for the different types of sources in the literature sample, revealed, with 95\% confidence, that extragalactic and low-mass Class 0 protostellar outflows follow the same correlation, thus suggesting the presence of a common underlying launch mechanism.
    \item An order-of-magnitude calculation, comparing the total outflow force contribution of low- and high-mass protostars, suggests that low-mass protostars dominate the total protostellar contribution in extragalactic starburst outflows.
\end{itemize}

The results presented in this paper allow for some interesting conclusions, but they also highlight the need for additional research and observations on the matter.
More specifically, additional and more accurate observations are certainly required in order to constrain the observed protostellar correlations. Constraining these correlations and increasing the number of sources from regions of clustered star formation would solidify the conclusions presented in this work.
This need is even more apparent in the case of intermediate-mass sources, required to properly connect the low- and high-mass protostellar samples, as such observations would allow for a better understanding of the observed differences between the various subsamples and the mechanisms behind it thus offering a more clear view of the overall behavior of outflows on all scales.

At this point though the data clearly point toward a correlation between the physical parameters of Galactic and extragalactic outflows strongly supporting the existence of a common launching mechanism for both types of outflows. As such, this paves the way for developing a universal outflow launch model.

\begin{acknowledgements}
We would like to thank the anonymous referee for suggestions that greatly helped in clarifying the presentation of the results, and for constructive comments on the manuscript in general. We would also like to thank Susanne Aalto for discussing extragalactic outflows, and Sarel van der Walt for assistance with the SubMillimeter Array data. The Submillimeter Array is a joint project between the Smithsonian Astrophysical Observatory and the Academia Sinica Institute of Astronomy and Astrophysics and is funded by the Smithsonian Institution and the Academia Sinica. We acknowledge and thank the staff of the SMA for their assistance and continued support. The authors wish to recognize and acknowledge the very significant cultural role and reverence that the summit of Mauna Kea has always had within the indigenous Hawaiian community. We are most fortunate to have had the opportunity to conduct observations from this mountain. The research of IMS and LEK is supported by a research grant (19127) from VILLUM FONDEN. 
\end{acknowledgements}

\bibliographystyle{aa}

\bibliography{bibliography.bib}

\begin{appendix}

\section{Outflow mass derivation}
\label{sec:masscalc}

In order to measure the mass of the outflow, it is standard to assume that the emission from the outflow is optically thin \citep{dunham2014b} and that the CO level populations in the outflowing gas are in local thermodynamic equilibrium (LTE) \citep[see Appendix A in ][]{maud2015I}). In this case, the column density of the upper level of the observed transition is:

\begin{equation}
    N_{\text{u}} = \frac{8 \pi k_\text{B}}{h c^3} \frac{\nu^2 \int T_\text{B} \text{d}v}{A_\text{ul}} ,
    \label{eq:eq4}
\end{equation}

where $k_\text{B}$ is Boltzmann's constant, $h$ is Planck's constant, $c$ the speed of light, $\nu$ the frequency of the emission line, and $T_\text{B}$ the brightness temperature, which is connected to the observed intensity as $\displaystyle I_\nu = 2kT_\text{B} \nu^2 / c^2 $, and is integrated over the velocity range covered by the outflow.
The total column density is then given by: 

\begin{equation}
    N_\text{CO} = \frac{Q(T_\text{ex})}{g_\text{u}} \text{e}^{(E_\text{u}/{k_\text{B}T_\text{ex}})} N_\text{u} ,
    \label{eq:eq5}
\end{equation}

with $Q{(T_\text{ex})}$ the partition function, $g_\text{u}$ the degeneracy of the upper level, $E_\text{u}$ the energy of the upper level, and $T_\text{ex}$ the excitation temperature of the observed gas. 
Multiplying with the abundance ratio of CO to H$_2$, $\displaystyle \left[ \text{H}_2 / \text{CO}\right]$, gives the total column density for H$_2$ and then the total mass can be calculated as:

\begin{equation}
    M = \mu m_\text{H} N_{\text{H}_2}A ,
    \label{eq:eq6}
\end{equation}

where $\mu$ is the mean molecular weight, $m_\text{H}$ the hydrogen mass, and $A$ is the observed area.
By combining Eq. \ref{eq:eq4}, \ref{eq:eq5}, \ref{eq:eq6} we get that the mass is:

\begin{equation}
    M = \mu m_\text{H} A \left[ \frac{\text{H}_2}{\text{CO}}\right] \frac{Q{(T_\text{ex})}}{g_\text{u}} \text{e}^{(E_\text{u}/{k_\text{B}T_\text{ex}})} \frac{8 \pi k_\text{B}\nu^2}{h c^3 A_\text{ul}} \int T_\text{B} \text{d}v \ .
\end{equation}

All the constants can be gathered in a conversion factor $K$ used to convert integrated intensity into mass. $K$ is then defined as:

\begin{equation}
    K = \mu m_\text{H} A \left[ \frac{\text{H}_2}{\text{CO}}\right] \frac{Q{(T_\text{ex})}}{g_\text{u}} \text{e}^{(E_\text{u}/{k_\text{B}T_\text{ex}})} \frac{8 \pi k_\text{B}\nu^2}{h c^3 A_\text{ul}} .
    \label{eq:kappa}
\end{equation}

In this work, for calculating $K$, a mean molecular weight $\mu = 2.8$ was used in order to also take into account He \citep{kauffmann2008}, a CO to H$_2$ abundance ratio of $\displaystyle \left[ \text{H}_2 / \text{CO}\right] = 1.2 \times 10^4$ \citep{frerking1982} and an excitation temperature $T_\text{ex} = 50$ K, as used in \citet{vdm2013,yildiz2012}. The values used for the partition function is $Q = 18.5$ (corresponding to $T_{\rm ex}$ of 50 K), the upper level degeneracy $g_\text{u}= 7$, the upper level energy $E_\text{u} = 33.2$ K, and the Einstein A-coefficient $A_\text{ul} = 10^{-5.6}$ s$^{-1}$; these values are taken from Splatalogue\footnote{\url{https://splatalogue.online/advanced1.php}} and correspond to the CO $J$ = 3--2 transition examined in this work. The selected temperature is found to not significantly affect the results as, for example, using $T_\text{ex} = 100$ K only leads to an increase of the final outflow forces by a factor of $\sim$1.4
Lastly, $A$ is now the physical area covered by one pixel and is calculated assuming a distance of 1.4 kpc, except for S26 for which a distance of 3.3 kpc is used. The distances used are consistent with those used in \citet{pitts2021} and are based on measurements from \citet{rygl2012}.

\section{The separation method}
\label{sec:separationapp}
In the separation method, the outflow force is calculated by:
\begin{equation}
    F_\text{CO} = c_3 \times \frac{\displaystyle K \left( \sum_\text{j} \left[ \int_{v_\text{in}}^{v_\text{out,j}}T(v')v'\text{d}v' \right]_\text{j}\right) v_\text{max}}{\displaystyle R_\text{lobe}} .
\end{equation}
In this  equation, $c_3$ is a correction factor for the inclination of the outflow and can take the values shown in Table \ref{table:inclinations} according to the inclination angle of the outflow. Due to the large uncertainties in determining the actual inclination angle of the outflows from their morphology $c_3$ takes four different values representing the inclination angle ranges of 0$\degr$ -- 30$\degr$, 30$\degr$ -- 50$\degr$, 50$\degr$ -- 70$\degr$, and 70$\degr$ -- 90$\degr$ respectively.
For this work, the correction factors are calculated, following the method presented in \citet{vdm2013}, by taking the mean of the $t_{v_\text{max}}/$ True age values for mass density contrasts of 0.1 and 1 presented in Table 6 in \citet{downes2007} for velocities of 7 -- 40 km s$^{-1}$. Contrary to \citet{vdm2013}, who used the table values for a velocity range of 2 -- 40 km s$^{-1}$, the 7 -- 40 km s$^{-1}$ velocity range is deemed more appropriate here, where most of the emission at low velocities is missing.
In this work, the outflow forces are calculated using all four values of $c_3$ as to both make up for the uncertainty inherent in determining the inclination angle of an outflow but also to examine the effect these uncertainties actually have on the final result.

Next, the sum of the integrated intensities is performed over all pixels contributing to the outflow to get the total velocity-weighted integrated intensity of the outflow.
In order to properly determine the pixels contributing to the outflow, we isolate the outflow area, aiming to avoid including unrelated structures in the measurements, and then sum up all pixels with an integrated intensity higher than the 3$\sigma_\text{rms}$ ($\sigma_\text{red}$ and $\sigma_\text{blue}$ for the red and blue-shifted lobe respectively) level used in the contour maps of the outflows (Fig. \ref{fig:outflows2}) while the velocity-weighted data cube is created by multiplying each channel with the corresponding velocity.

The integration itself is performed for each pixel separately, with velocity limits $v_\text{in}$ and $v_\text{out,j}$. The lower velocity limit $v_\text{in}$ is the same for all pixels but $v_\text{out,j}$ is calculated individually for each pixel as the highest velocity at which the intensity remains above the 1 $\sigma_\text{rms}$ level.
For the lower limit $v_\text{in}$, the inner velocities $v_\text{b,in}$ and $v_\text{r,in}$ shown in table \ref{table:velrange} are used, while the upper limit $v_\text{out,j}$ is calculated as described previously using as $\sigma_\text{rms}$ limit the single channel rms noise of each source, presented in Table \ref{table:beamsizes}.

Then, $v_\text{max}$ is the maximum outflow velocity, used for the calculation of $t_\text{dyn}$, and is calculated as the maximum (in absolute value for the blue-shifted lobes) of the $v_\text{out,j}$ for each outflow lobe.
Finally, $R_\text{lobe}$, the projected extent of the outflow lobes is measured directly from the contour maps.
The values of $v_\text{max}$ and $R_\text{lobe}$ measured for our sample are given in table \ref{table:maxvel}.

\section{Outflow forces for different inclination angles and calculation methods}
\label{sec:outflowforcesApp}
Outflow forces as calculated using the Separation, $v_\text{max}$ and $\langle v \rangle$ methods for all four different possible inclination angles.
\begin{table*}[t]
\caption{$F_\text{CO}$ calculated with the separation method for the red-shifted lobe, the blue-shifted lobe and the whole outflow.} 
\label{table:separation}
\centering 
\begin{tabular}{l c c c c c c c c c c c c} 
\hline\hline 
 \multicolumn{1}{c|}{$i$ (\degr)}& \multicolumn{3}{c|}{10} & \multicolumn{3}{c|}{30} & \multicolumn{3}{c|}{50} & \multicolumn{3}{c}{70} \\ 
\hline
Source & $F_\text{red}$ & $F_\text{blue}$ & $F_\text{total}$ & $F_\text{red}$ & $F_\text{blue}$ & $F_\text{total}$ & $F_\text{red}$ & $F_\text{blue}$ & $F_\text{total}$ & $F_\text{red}$ & $F_\text{blue}$ & $F_\text{total}$ \\
\hline
N12 & \phantom{1}1.1 & \phantom{1}2.2 & \phantom{1}3.3  & \phantom{1}2.4 & \phantom{1}4.8 & \phantom{1}7.2 & \phantom{1}4.5& 8.8 & 13.3 & 7.2 & 13.9 & 21.1 \\
N30 & 73.7 & 13.5 & 87.2 & 159.8\phantom{1} & 29.3 & 189.1\phantom{1} & 294.9\phantom{1} & 54.1& 349.0\phantom{1} & 467.0\phantom{1} & 85.6\phantom{1} & 552.6\phantom{1}\\ 
N38 & \phantom{1}0.3 & \phantom{1}0.9 & \phantom{1}1.2 & \phantom{1}0.6 & \phantom{1}2.0 & \phantom{1}2.6 & \phantom{1}1.0 & \phantom{1}3.7 & \phantom{1}4.7 & \phantom{1}1.6 & \phantom{1}5.9 & 7.5 \\ 
N48 & \phantom{1}1.4 & \phantom{1}2.7 & \phantom{1}4.1 & \phantom{1}3.0 & \phantom{1}5.9 & 8.9 & \phantom{1}5.5 & 10.9 & 16.4 & 8.8 & 17.2 & 26.0 \\ 
N51 & \phantom{1}0.5 & \phantom{1}2.8 & \phantom{1}3.3 & \phantom{1}1.0 & \phantom{1}6.1 & 7.1 & \phantom{1}1.9 & 11.3 & 13.2 & \phantom{1}3.0 & 17.9 & 20.9 \\ 
N53 & \phantom{1}0.3 & 7.4 & 7.7 & \phantom{1}0.7 & 16.1 & 16.8 & \phantom{1}1.3 & 29.7 & 31.0 & \phantom{1}2.0 & 47.1 & 49.1 \\ 
N54 & \phantom{1}1.7 & \phantom{1}0.9 & \phantom{1}2.6 & \phantom{1}3.8 & \phantom{1}1.9 & \phantom{1}5.7 & \phantom{1}7.0 & \phantom{1}3.5 & 10.5 & 11.0 & \phantom{1}5.6 & 16.6 \\ 
N63 & 8.2 & 11.7 & 19.9 & 17.9 & 25.3 & 43.2 & 33.0 & 46.8 & 79.8 & 52.3 & 74.1 & 126.4\phantom{1} \\ 
S8  & \phantom{1}2.2 & \phantom{1}4.8 & 7.0 & \phantom{1}4.9 & 10.4 & 15.3 & 9.0 & 19.3 & 28.3 & 14.2 & 30.6 & 44.8 \\ 
S26 & 15.0 & 27.2 & 42.2 & 32.5 & 58.9 & 91.4\phantom{1} & 60.0 & 108.7\phantom{1} & 168.7\phantom{1} & 95.0 & 172.1\phantom{1} & 267.1\phantom{1} \\ 
\hline 
\end{tabular}
\tablefoot{ All results are shown in units of 10$^{-5}$ M$_\odot$ yr$^{-1}$ km s$^{-1}$ .}
\end{table*}

\begin{table*}
\caption{$F_\text{CO}$ calculated with the $v_\text{max}$ method for the red-shifted lobe, the blue-shifted lobe and the whole outflow.} 
\label{table:vmax} 
\centering 
\begin{tabular}{l c c c c c c c c c c c c}  
\hline\hline 
 \multicolumn{1}{c|}{$i(\degr)$}& \multicolumn{3}{c|}{10} & \multicolumn{3}{c|}{30} & \multicolumn{3}{c|}{50} & \multicolumn{3}{c}{70} \\ 
\hline
Source & $F_\text{red}$ & $F_\text{blue}$ & $F_\text{total}$ & $F_\text{red}$ & $F_\text{blue}$ & $F_\text{total}$ & $F_\text{red}$ & $F_\text{blue}$ & $F_\text{total}$ & $F_\text{red}$ & $F_\text{blue}$ & $F_\text{total}$ \\
\hline
N12 & \phantom{1}1.5 & \phantom{1}2.1 & \phantom{1}3.6 & \phantom{1}2.4 & \phantom{1}3.4 & \phantom{1}5.8 & \phantom{1}2.4 & \phantom{1}3.4 & \phantom{1}5.8 & \phantom{1}5.9 & \phantom{1}8.2 & 14.1  \\ 
N30 & 81.7\phantom{1} & 9.7 & 91.4\phantom{1} & 131.3\phantom{1} & 15.6 & 146.9\phantom{1} & 131.3\phantom{1} & 15.6 & 146.9\phantom{1} & 320.9\phantom{1} & 38.2 & 359.1\phantom{1} \\ 
N38 & \phantom{1}0.2 & \phantom{1}1.1 & \phantom{1}1.3 & \phantom{1}0.4 & \phantom{1}1.8 & \phantom{1}2.2 & \phantom{1}0.4 & \phantom{1}1.8 & \phantom{1}2.2 & \phantom{1}1.0 & \phantom{1}4.3 & \phantom{1}5.3 \\ 
N48 & \phantom{1}1.1 & \phantom{1}6.3 & \phantom{1}7.4 & \phantom{1}1.8 & 10.2 & 12.0 & \phantom{1}1.8 & 10.2 & 12.0 & \phantom{1}4.4 & 24.9 & 29.3 \\ 
N51 & \phantom{1}0.4 & \phantom{1}2.1 & \phantom{1}2.5 & \phantom{1}0.7 & \phantom{1}3.3 & \phantom{1}4.0 & \phantom{1}0.7 & \phantom{1}3.3 & \phantom{1}4.0 & \phantom{1}1.7 & 8.1 & 9.8 \\ 
N53 & \phantom{1}0.2 & \phantom{1}6.8 & \phantom{1}7.0 & \phantom{1}0.3 & 10.9 & 11.2 & \phantom{1}0.3 & 10.9 & 11.2 & \phantom{1}0.8 & 26.6 & 27.4 \\ 
N54 & \phantom{1}3.6 & \phantom{1}0.7 & \phantom{1}4.3 & \phantom{1}5.8 & \phantom{1}1.1 & \phantom{1}6.9 & \phantom{1}5.8 & \phantom{1}1.1 & \phantom{1}6.9 & 14.2 & \phantom{1}2.7 & 16.9 \\ 
N63 & 8.6 & 12.3 & 20.9 & 13.9 & 19.7 & 33.6 & 13.9 & 19.7 & 33.6 & 33.9 & 48.2 & 82.1 \\ 
S8  & \phantom{1}3.8 & \phantom{1}4.9 & 8.7 & \phantom{1}6.0 & 7.9 & 13.9 & \phantom{1}6.0 & 7.9 & 13.9 & 14.8 & 19.3 & 34.1 \\ 
S26 & 18.2 & 18.4 & 36.6 & 29.2 & 29.6 & 58.8 & 29.2 & 29.6 & 58.8 & 71.4 & 72.4 & 143.8\phantom{1} \\ 
\hline 
\end{tabular}
\tablefoot{ All results are shown in units of 10$^{-5}$ M$_\odot$ yr$^{-1}$ km s$^{-1}$ .}
\end{table*}

\begin{table*}
\caption{$F_\text{CO}$ calculated with the $\langle v \rangle$ method for the red-shifted lobe, the blue-shifted lobe and the whole outflow.} 
\label{table:vmean} 
\centering 
\begin{tabular}{l c c c c c c c c c c c c} 
\hline\hline 
 \multicolumn{1}{c|}{$i(\degr)$}& \multicolumn{3}{c|}{10} & \multicolumn{3}{c|}{30} & \multicolumn{3}{c|}{50} & \multicolumn{3}{c}{70} \\ 
\hline
Source & $F_\text{red}$ & $F_\text{blue}$ & $F_\text{total}$ & $F_\text{red}$ & $F_\text{blue}$ & $F_\text{total}$ & $F_\text{red}$ & $F_\text{blue}$ & $F_\text{total}$ & $F_\text{red}$ & $F_\text{blue}$ & $F_\text{total}$ \\
\hline
N12 & \phantom{1}1.1 & \phantom{1}2.9 & \phantom{1}4.0 & \phantom{1}2.4 & \phantom{1}6.5 & 8.9 & \phantom{1}4.2 & 11.4 & 15.6 & 9.4 & 25.2 & 34.6 \\ 
N30 & 82.8 & 23.3 & 106.1\phantom{1} & 186.4\phantom{1} & 52.5 & 238.9\phantom{1} & 326.1\phantom{1} & 91.9 & 418.0\phantom{1} & 724.8\phantom{1} & 204.2\phantom{1} & 929.0\phantom{1} \\ 
N38 & \phantom{1}0.3 & \phantom{1}1.0 & \phantom{1}1.3 & \phantom{1}0.7 & \phantom{1}2.2 & \phantom{1}2.9 & \phantom{1}1.3 & \phantom{1}3.9 & \phantom{1}5.2 & \phantom{1}2.9 & 8.6 & 11.5 \\ 
N48 & \phantom{1}2.1 & \phantom{1}1.4 & \phantom{1}3.5 & \phantom{1}4.8 & \phantom{1}3.2 & 8.0 & 8.4 & \phantom{1}5.7 & 14.1 & 18.7 & 12.6 & 31.3 \\ 
N51 & \phantom{1}0.6 & \phantom{1}4.8 & \phantom{1}5.4 & \phantom{1}1.4 & 10.8 & 12.2 & \phantom{1}2.4 & 19.0 & 21.4 & \phantom{1}5.4 & 42.2 & 47.6 \\ 
N53 & \phantom{1}0.6 & 10.2 & 10.8 & \phantom{1}1.4 & 22.9 & 24.3 & \phantom{1}2.4 & 40.0 & 42.4 & \phantom{1}5.3 & 88.9\phantom{1} & 94.2\phantom{1} \\ 
N54 & \phantom{1}1.0 & \phantom{1}1.4 & \phantom{1}2.4 & \phantom{1}2.4 & \phantom{1}3.2 & \phantom{1}5.6 & \phantom{1}4.1 & \phantom{1}5.6 & 9.7 & 9.2 & 12.3 & 21.5 \\ 
N63 & 9.8 & 13.9 & 23.7 & 22.1 & 31.2 & 53.3 & 38.6 & 54.7 & 93.3 & 85.8 & 121.5\phantom{1} & 207.3\phantom{1} \\ 
S8  & \phantom{1}1.7 & \phantom{1}5.9 & 7.6 & \phantom{1}3.7 & 13.3 & 17.0 & \phantom{1}6.5 & 23.3 & 29.8 & 14.5 & 51.7 & 66.2 \\ 
S26 & 15.4 & 49.8 & 65.2 & 34.7 & 112.1\phantom{1} & 146.8\phantom{1} & 60.7 & 196.1\phantom{1} & 256.8\phantom{1} & 134.8\phantom{1} & 435.9\phantom{1} & 570.7\phantom{1} \\ 
\hline 
\end{tabular}
\tablefoot{ All results are shown in units of 10$^{-5}$ M$_\odot$ yr$^{-1}$ km s$^{-1}$ .}
\end{table*}
\newpage
\FloatBarrier
\section{Outflow force - bolometric luminosity correlation best fits}
\label{sec:bestfitapp}
$F_\text{CO}$ - $M_\text{env}$ correlation plots, with the calculated best-fits of the various subsamples plotted with dashed lines.
\begin{figure*}[h!]
    \centering
    \includegraphics[width=0.45\linewidth]{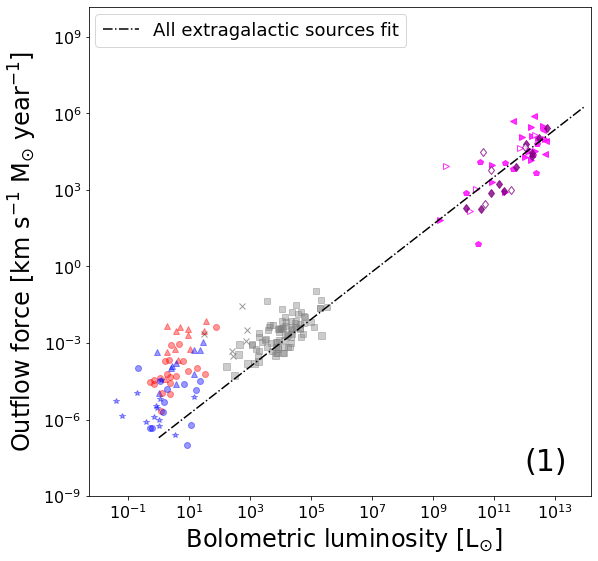}
    \includegraphics[width=0.45\linewidth]{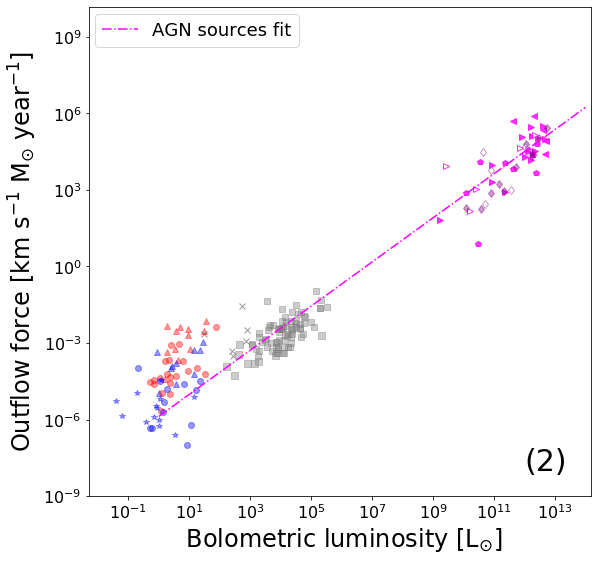}
    \includegraphics[width=0.45\linewidth]{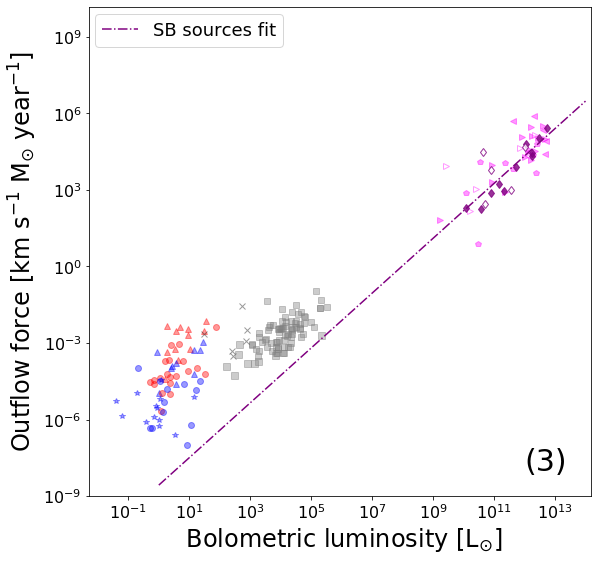}
    \includegraphics[width=0.45\linewidth]{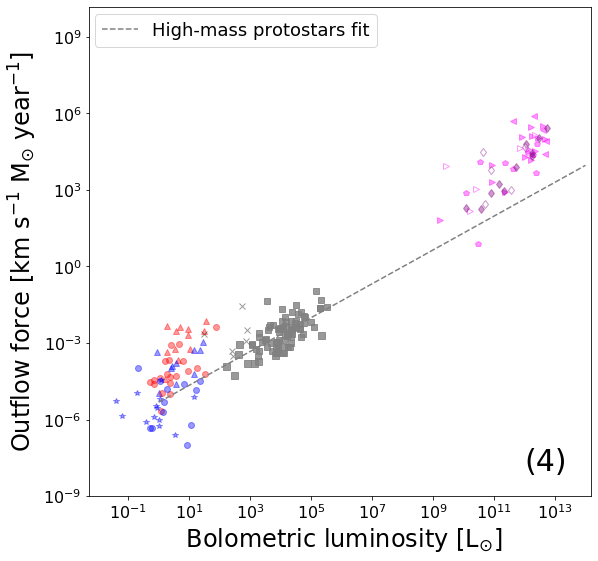}
    \includegraphics[width=0.45\linewidth]{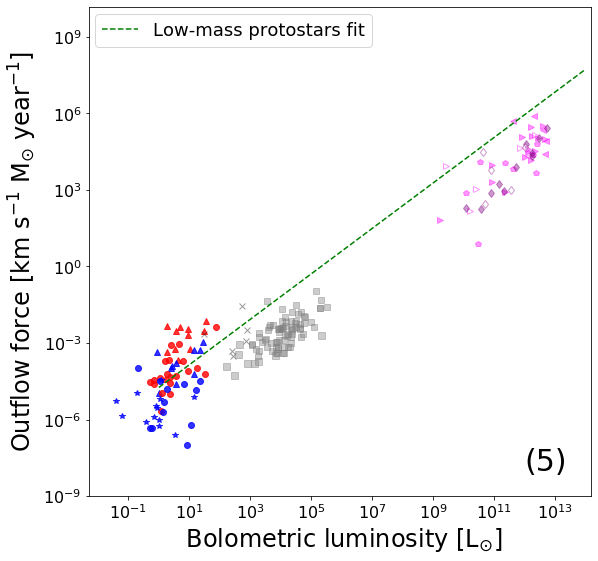}
    \includegraphics[width=0.45\linewidth]{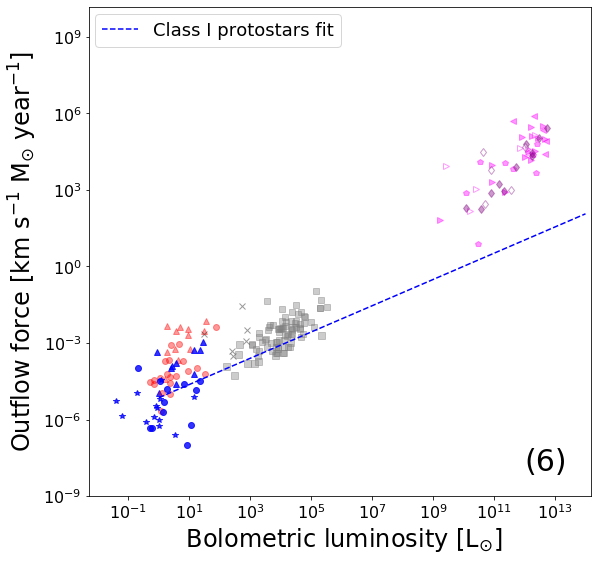}
    \caption{Outflow forces over bolometric luminosities for a sample of low-mass protostars, a sample of high-mass protostars and extragalactic sources. In each plot, the the sample, from which the plotted best-fit of the $F_\text{CO}$ - $L_\text{bol}$ was taken, is highlighted. (1) Best-fit to all extragalactic sources. (2) Best-fit to AGN extragalactic sources. (3) Best-fit to starburst extragalactic sources. (4) Best-fit to high-mass protostellar sources. (5) Best-fit to all low-mass protostellar sources. (6) Best-fit to Class I low-mass protostellar sources. (7) Best-fit to Class 0 low-mass protostellar sources. }
    \label{fig:fits}
    \end{figure*}

\begin{figure*}[h]  
    \centering
    \includegraphics[width=0.45\linewidth]{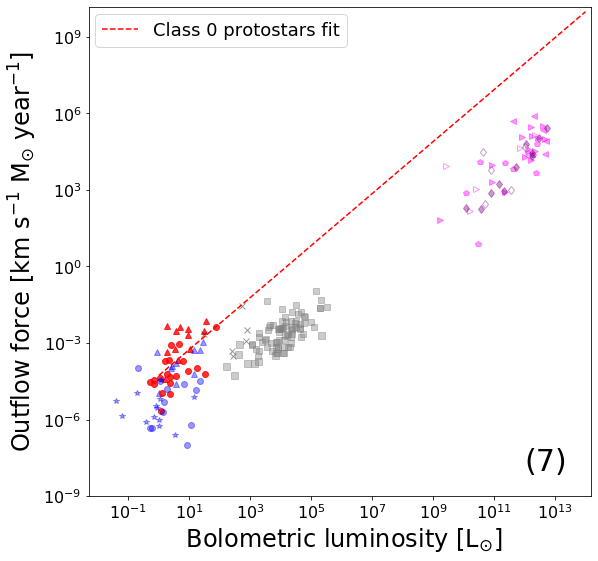}
    \caption{Fig. \ref{fig:fits} continued.}
    \label{fig:fits_no}
\end{figure*}

\newpage
\clearpage
\section{Spectra for determining integration velocity ranges.}
\label{sec:appendixA}

All the spectra used for determining the velocity ranges used to create the CO integrated emission maps of the Cygnus-X sources presented in Section \ref{sec:maps}.

\subsection{N12}

The spectra of N12 used to determine the velocity ranges for the creation of the integrated CO emission maps.

\begin{figure*}[htb]
\centering
\includegraphics[width=0.4\linewidth]{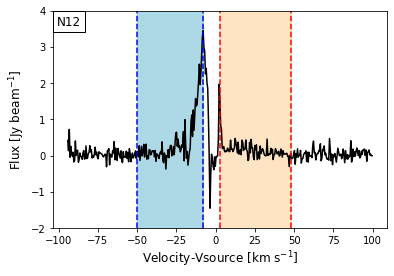}
\caption{Spectrum of N12 from the location of peak continuum emission. } 
\end{figure*}

\begin{figure*}[htb]
\centering
\includegraphics[width=0.4\linewidth]{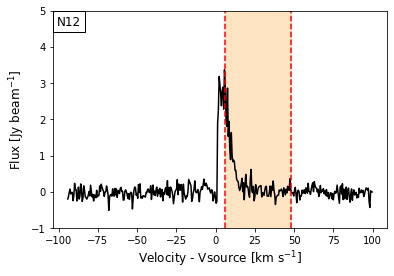}
\includegraphics[width=0.4\linewidth]{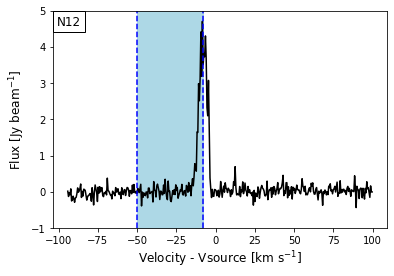}
\caption{[Left]: Spectrum of N12 from the location of peak red-shifted emission. [Right]: Spectrum of N12 from the location of peak blue-shifted emission. }
\end{figure*}

\begin{figure*}[htb]
\centering
\includegraphics[width=0.4\linewidth]{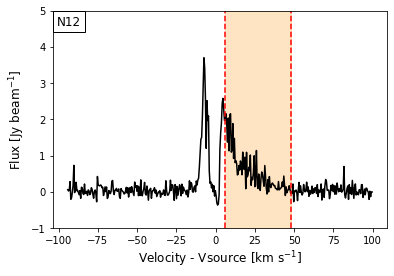}
\includegraphics[width=0.4\linewidth]{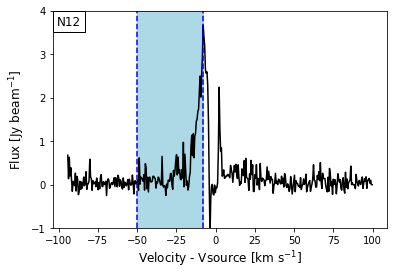}
\caption{[Left]: Spectrum of N12 from the location of the peak of the integrated red-shifted emission. [Right]: Spectrum of N12 from the location of the peak of the integrated blue-shifted emission.} 
\end{figure*}

\clearpage

\subsection{N30}

The spectra of N30 used to determine the velocity ranges for the creation of the integrated CO emission maps.
\begin{figure*}[h]
\centering
\includegraphics[width=0.4\linewidth]{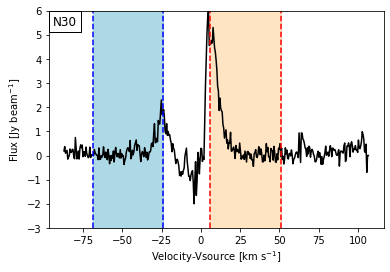}
\caption{Spectrum of N30 from the location of peak continuum emission. } 
\end{figure*}

\begin{figure*}[h!]
\centering
\includegraphics[width=0.4\linewidth]{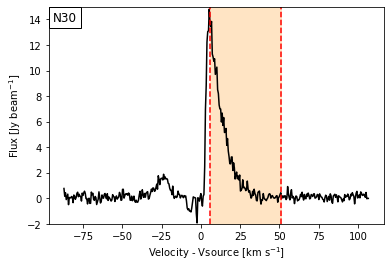}
\includegraphics[width=0.4\linewidth]{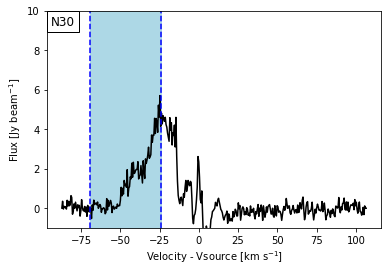}
\caption{[Left]: Spectrum of N30 from the location of peak red-shifted emission. [Right]: Spectrum of N30 from the location of peak blue-shifted emission. }
\end{figure*}

\begin{figure*}[h]
\centering
\includegraphics[width=0.4\linewidth]{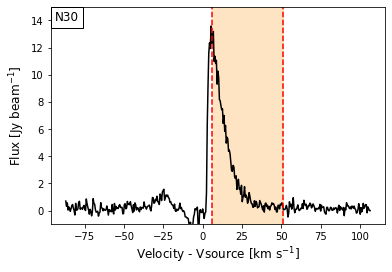}
\includegraphics[width=0.4\linewidth]{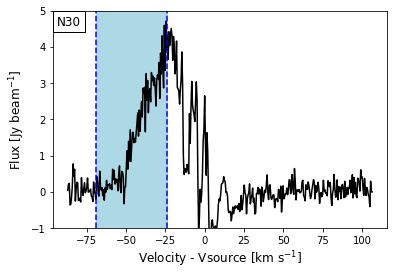}
\caption{[Left]: Spectrum of N30 from the location of the peak of the integrated red-shifted emission. [Right]: Spectrum of N30 from the location of the peak of the integrated blue-shifted emission.} 
\end{figure*}
\clearpage

\subsection{N38}
The spectra of N38 used to determine the velocity ranges for the creation of the integrated CO emission maps.
\begin{figure*}[h]
\centering
\includegraphics[width=0.4\linewidth]{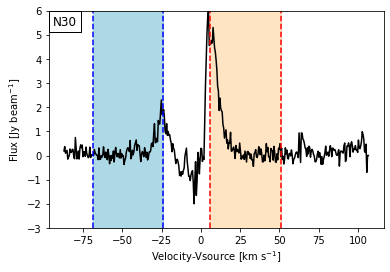}
\caption{Spectrum of N30 from the location of peak continuum emission.}
\end{figure*}

\begin{figure*}[h!]
\centering
\includegraphics[width=0.4\linewidth]{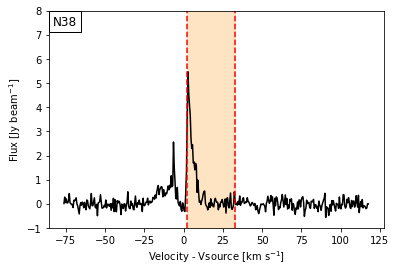}
\includegraphics[width=0.4\linewidth]{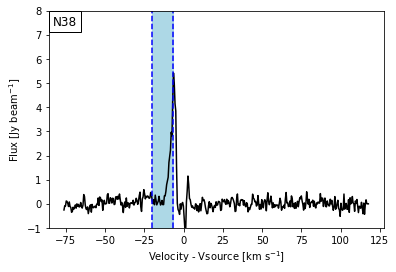}
\caption{[Left]: Spectrum of N38 from the location of peak red-shifted emission. [Right]: Spectrum of N38 from the location of peak blue-shifted emission. }
\end{figure*}

\begin{figure*}[h]
\centering
\includegraphics[width=0.4\linewidth]{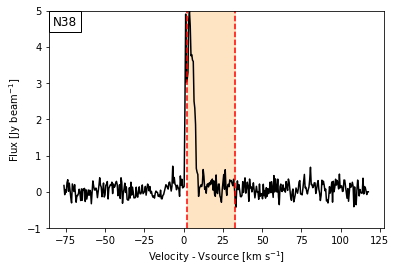}
\includegraphics[width=0.4\linewidth]{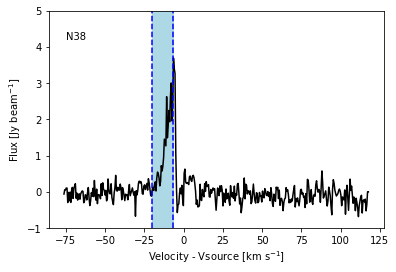}
\caption{[Left]: Spectrum of N38 from the location of the peak of the integrated red-shifted emission. [Right]: Spectrum of N38 from the location of the peak of the integrated blue-shifted emission.} 
\end{figure*}
\clearpage

\subsection{N48}
The spectra of N48 used to determine the velocity ranges for the creation of the integrated CO emission maps.
\begin{figure*}[h]
\centering
\includegraphics[width=0.4\linewidth]{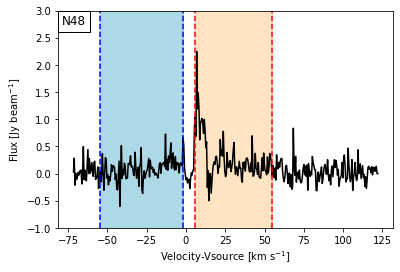}
\caption{Spectrum of N48 from the location of peak continuum emission.}
\end{figure*}

\begin{figure*}[h!]
\centering
\includegraphics[width=0.4\linewidth]{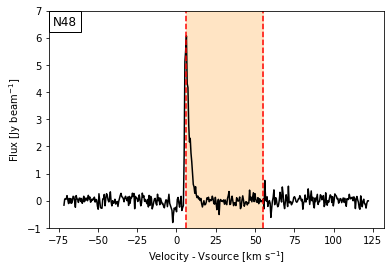}
\includegraphics[width=0.4\linewidth]{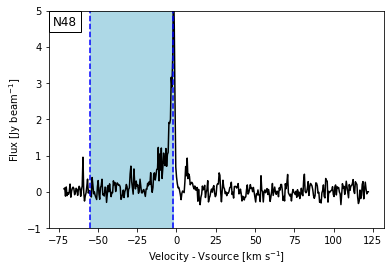}
\caption{[Left]: Spectrum of N48 from the location of peak red-shifted emission. [Right]: Spectrum of N48 from the location of peak blue-shifted emission. }
\end{figure*}

\begin{figure*}[h]
\centering
\includegraphics[width=0.4\linewidth]{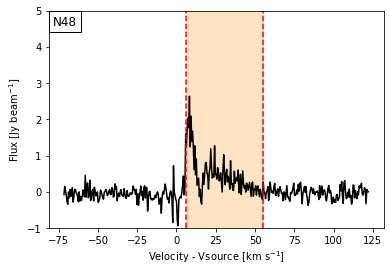}
\includegraphics[width=0.4\linewidth]{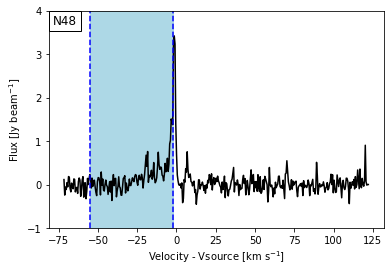}
\caption{[Left]: Spectrum of N48 from the location of the peak of the integrated red-shifted emission. [Right]: Spectrum of N48 from the location of the peak of the integrated blue-shifted emission.} 
\end{figure*}
\clearpage

\subsection{N51}
The spectra of N51 used to determine the velocity ranges for the creation of the integrated CO emission maps.
\begin{figure*}[h]
\centering
\includegraphics[width=0.4\linewidth]{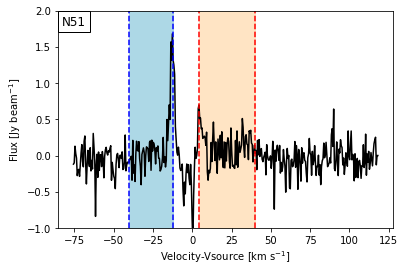}
\caption{Spectrum of N51 from the location of peak continuum emission.} 
\end{figure*}

\begin{figure*}[h!]
\centering
\includegraphics[width=0.4\linewidth]{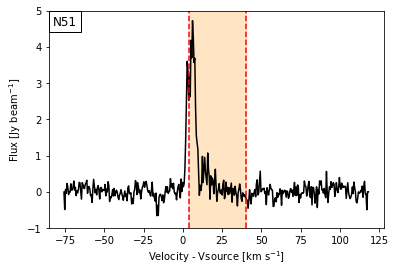}
\includegraphics[width=0.4\linewidth]{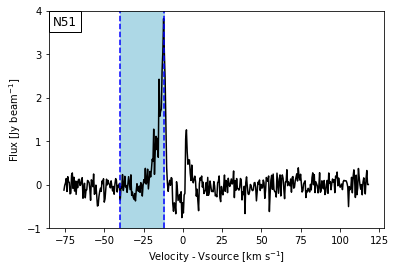}
\caption{[Left]: Spectrum of N51 from the location of peak red-shifted emission. [Right]: Spectrum of N51 from the location of peak blue-shifted emission. }
\end{figure*}

\begin{figure*}[h]
\centering

\caption{} 
\end{figure*}

\begin{figure*}[h]
\centering
\includegraphics[width=0.4\linewidth]{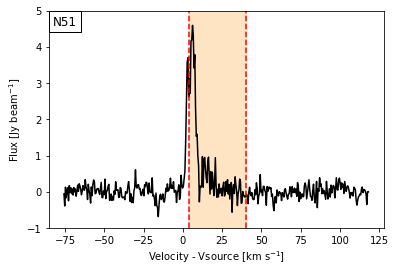}
\includegraphics[width=0.4\linewidth]{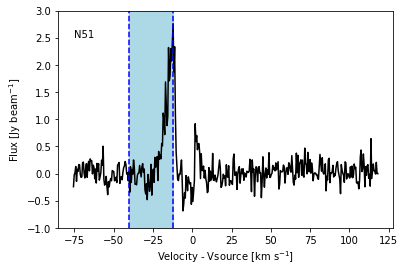}
\caption{[Left]: Spectrum of N51 from the location of the peak of the integrated red-shifted emission. [Right]:Spectrum of N51 from the location of the peak of the integrated blue-shifted emission.} 
\end{figure*}
\clearpage

\subsection{N53}
The spectra of N53 used to determine the velocity ranges for the creation of the integrated CO emission maps.
\begin{figure*}[h]
\centering
\includegraphics[width=0.4\linewidth]{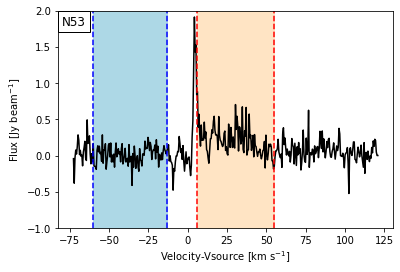}
\caption{Spectrum of N53 from the location of peak continuum emission.}
\end{figure*}

\begin{figure*}[h!]
\centering
\includegraphics[width=0.4\linewidth]{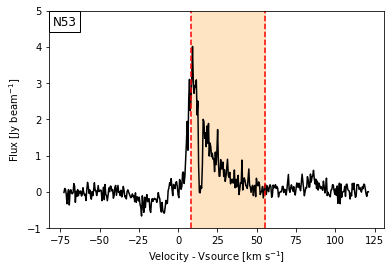}
\includegraphics[width=0.4\linewidth]{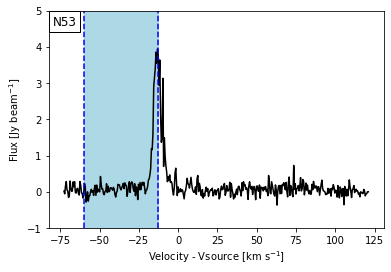}
\caption{[Left]: Spectrum of N53 from the location of peak red-shifted emission. [Right]: Spectrum of N53 from the location of peak blue-shifted emission. }
\end{figure*}

\begin{figure*}[h]
\centering
\includegraphics[width=0.4\linewidth]{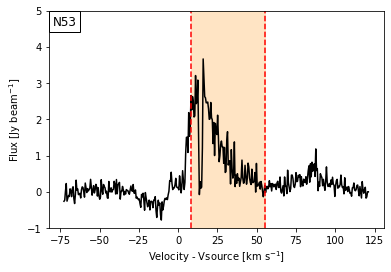}
\includegraphics[width=0.4\linewidth]{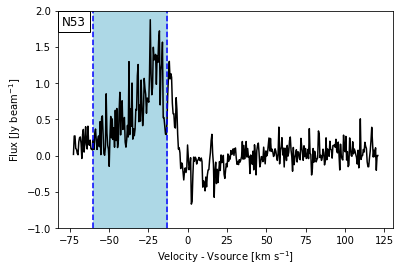}
\caption{[Left]: Spectrum of N53 from the location of the peak of the integrated red-shifted emission. [Right]: Spectrum of N53 from the location of the peak of the integrated blue-shifted emission.} 
\end{figure*}
\clearpage

\subsection{N54}
The spectra of N54 used to determine the velocity ranges for the creation of the integrated CO emission maps.
\begin{figure*}[h]
\centering
\includegraphics[width=0.4\linewidth]{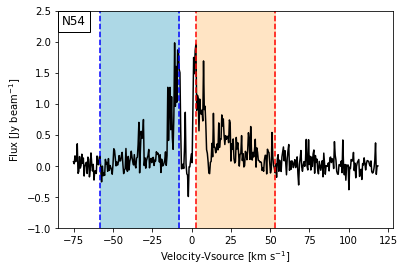}
\caption{Spectrum of N54 from the location of peak continuum emission.}
\end{figure*}

\begin{figure*}[h!]
\centering
\includegraphics[width=0.4\linewidth]{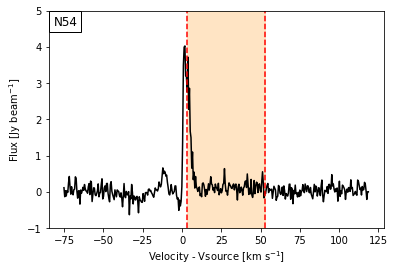}
\includegraphics[width=0.4\linewidth]{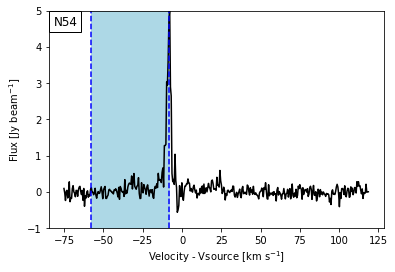}
\caption{[Left]: Spectrum of N54 from the location of peak red-shifted emission. [Right]: Spectrum of N54 from the location of peak blue-shifted emission. }
\end{figure*}

\begin{figure*}[h]
\centering
\includegraphics[width=0.4\linewidth]{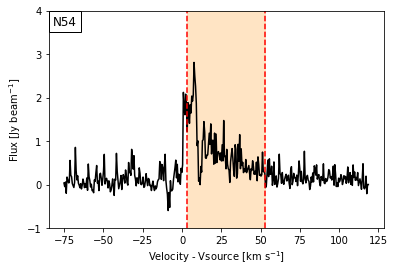}
\includegraphics[width=0.4\linewidth]{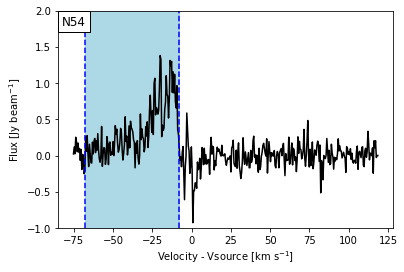}
\caption{[Left]: Spectrum of N54 from the location of the peak of the integrated red-shifted emission. [Right]: Spectrum of N54 from the location of the peak of the integrated blue-shifted emission.} 
\end{figure*}
\clearpage

\subsection{N63}
The spectra of N63 used to determine the velocity ranges for the creation of the integrated CO emission maps.
\begin{figure*}[h]
\centering
\includegraphics[width=0.4\linewidth]{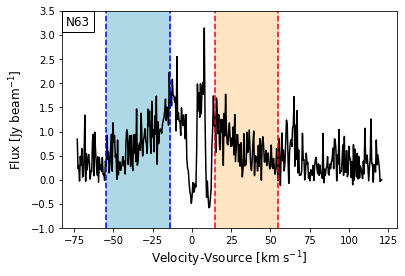}
\caption{Spectrum of N63 from the location of peak continuum emission.} 
\end{figure*}

\begin{figure*}[h!]
\centering
\includegraphics[width=0.4\linewidth]{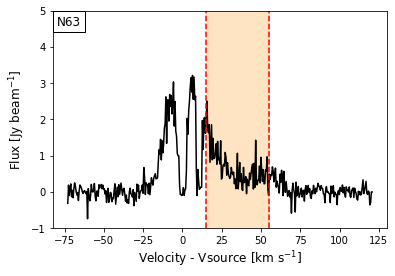}
\includegraphics[width=0.4\linewidth]{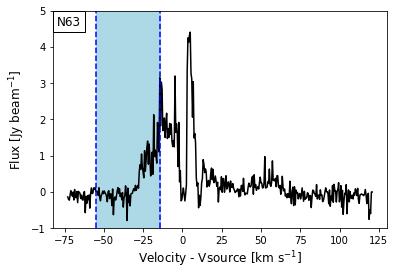}
\caption{[Left]: Spectrum of N63 from the location of peak red-shifted emission. [Right]: Spectrum of N63 from the location of peak blue-shifted emission. }
\end{figure*}

\begin{figure*}[h]
\centering
\includegraphics[width=0.4\linewidth]{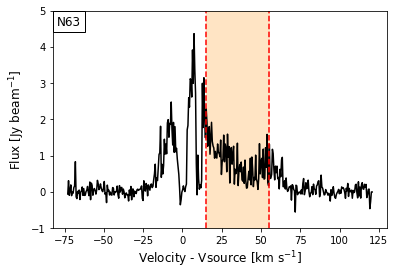}
\includegraphics[width=0.4\linewidth]{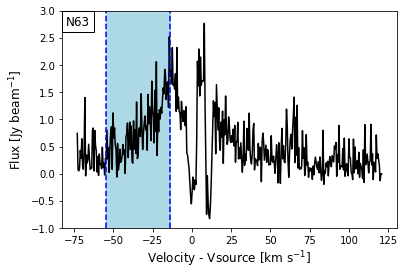}
\caption{[Left]: Spectrum of N63 from the location of the peak of the integrated red-shifted emission. [Right]: Spectrum of N63 from the location of the peak of the integrated blue-shifted emission.} 
\end{figure*}
\clearpage

\subsection{S8}
The spectra of S8 used to determine the velocity ranges for the creation of the integrated CO emission maps.
\begin{figure*}[h]
\centering
\includegraphics[width=0.4\linewidth]{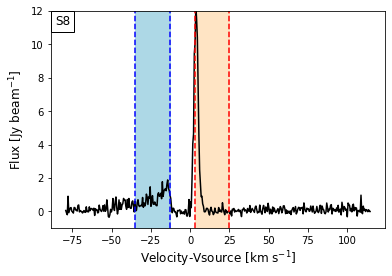}
\caption{Spectrum of S8 from the location of peak continuum emission. } 
\end{figure*}

\begin{figure*}[h!]
\centering
\includegraphics[width=0.4\linewidth]{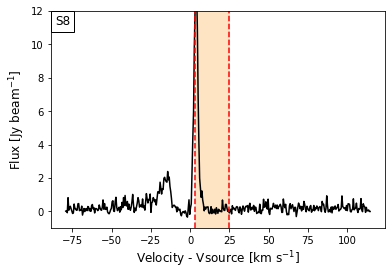}
\includegraphics[width=0.4\linewidth]{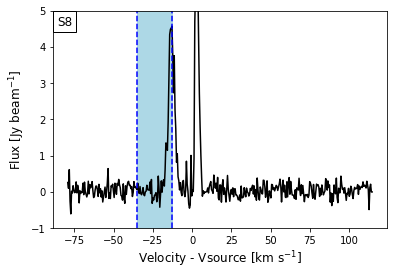}
\caption{[Left]: Spectrum of S8 from the location of peak red-shifted emission. [Right]: Spectrum of S8 from the location of peak blue-shifted emission. }
\end{figure*}

\begin{figure*}[h]
\centering
\includegraphics[width=0.4\linewidth]{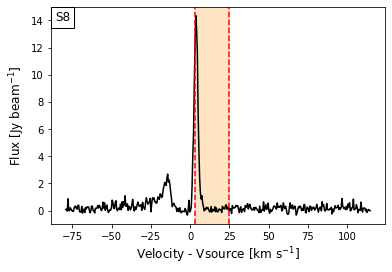}
\includegraphics[width=0.4\linewidth]{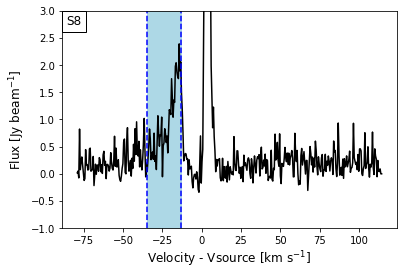}
\caption{[Left]: Spectrum of S8 from the location of the peak of the integrated red-shifted emission. [Right]:Spectrum of S8 from the location of the peak of the integrated blue-shifted emission.} 
\end{figure*}
\clearpage

\subsection{S26}
The spectra of S26 used to determine the velocity ranges for the creation of the integrated CO emission maps.
\begin{figure*}[h]
\centering
\includegraphics[width=0.4\linewidth]{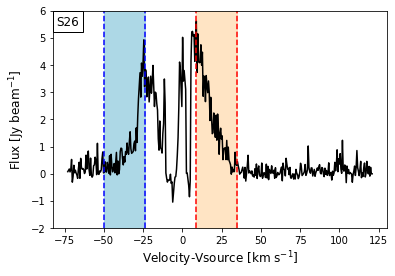}
\caption{Spectrum of S26 from the location of peak continuum emission. } 
\end{figure*}

\begin{figure*}[h!]
\centering
\includegraphics[width=0.4\linewidth]{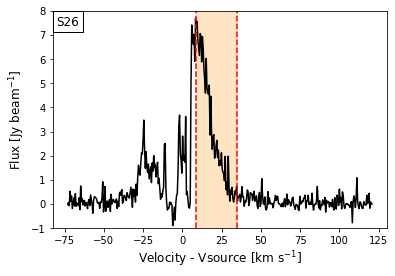}
\includegraphics[width=0.4\linewidth]{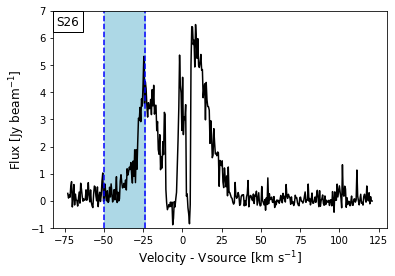}
\caption{[Left]: Spectrum of S26 from the location of peak red-shifted emission. [Right]: Spectrum of S26 from the location of peak blue-shifted emission. }
\end{figure*}

\begin{figure*}[h]
\centering
\includegraphics[width=0.4\linewidth]{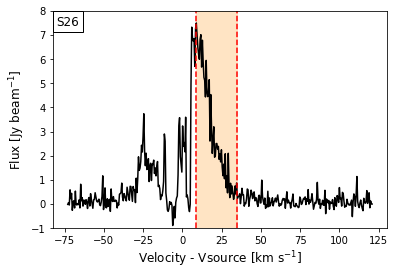}
\includegraphics[width=0.4\linewidth]{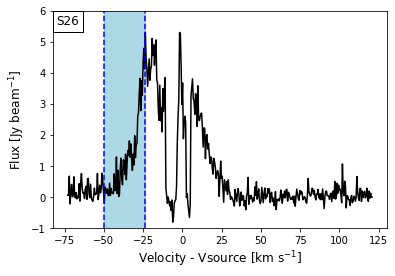}
\caption{[Left]: Spectrum of S26 from the location of the peak of the integrated red-shifted emission. [Right]: Spectrum of S26 from the location of the peak of the integrated blue-shifted emission.} 
\end{figure*}
\clearpage

\end{appendix}

\end{document}